\documentclass{ptptex} 
\usepackage{graphicx}
\usepackage{amsmath}

\newcommand{\del}{\partial} 

\newcommand{\beq}{\begin{equation}}
\newcommand{\eeq}{\end{equation}}
\newcommand{\bea}{\begin{eqnarray}}
\newcommand{\eea}{\end{eqnarray}}
\newcommand{\bsub}{\begin{subequations}}
\newcommand{\esub}{\end{subequations} \noindent}

\newcommand{\tr}{{\rm tr}}
\newcommand{\Seff}{S_{{\rm eff}}}
\newcommand{\Veff}{V_{{\rm eff}}}
\newcommand{\Repa}{{\rm Re}}
\newcommand{\Impa}{{\rm Im}}
 
\def\drvstar#1{\partial\kern-0.5pt\smash{\raise 4.5pt\hbox{$\ast$}}
               \kern-5.0pt_{#1}} 
\def\lvec#1{\setbox0=\hbox{$#1$}
    \setbox1=\hbox{$\scriptstyle\leftarrow$}
    #1\kern-\wd0\smash{
    \raise\ht0\hbox{$\raise1pt\hbox{$\scriptstyle\leftarrow$}$}}
    \kern-\wd1\kern\wd0} 
 
\def\ldrvstar#1{\lvec{\,\partial}\kern-0.5pt\smash{\raise 4.5pt\hbox{$\ast$}}
               \kern-5.0pt_{#1}} 
\begin{document}

\thispagestyle{empty}   
\vspace*{-15mm}   
\baselineskip 10pt   
\begin{flushright}   
\begin{tabular}{l}   
{May 2004} \\ 
{KUNS-1914 }\\
{RIKEN-TH-24}\\
{UTHEP-488} \\ 
{hep-th/0405076}   
\end{tabular}   
\end{flushright}   
\baselineskip 24pt   
\vglue 10mm   
\begin{center}   
{\Large\bf   
 Loops versus Matrices \\ 
 - The Nonperturbative Aspects of Noncritical String -
}  
\vspace{8mm}   
  
\baselineskip 18pt   
\renewcommand{\thefootnote}{
}
{   
\renewcommand{\thefootnote}{\alph{footnote}} 
 Masanori \textsc{Hanada},$^1$ \footnote{hana@gauge.scphys.kyoto-u.ac.jp}
 Masashi \textsc{Hayakawa},$^2$ \footnote{haya@riken.jp}
 Nobuyuki \textsc{Ishibashi},$^3$ \footnote{ishibash@het.ph.tsukuba.ac.jp}  \\
 Hikaru \textsc{Kawai},$^{1,2}$ \footnote{hkawai@gauge.scphys.kyoto-u.ac.jp}
Tsunehide \textsc{Kuroki},$^2$ \footnote{kuroki@riken.jp}
Yoshinori \textsc{Matsuo}$^1$
\footnote{ymatsuo@gauge.scphys.kyoto-u.ac.jp} \\
and Tsukasa \textsc{Tada}$^2$ \footnote{tada@riken.jp}
}   
\vspace{5mm}   

{\it  
$^1$ Department of Physics, Kyoto University, 
     Kyoto 606-8502, Japan 
\\ 
$^2$ Theoretical Physics Laboratory, RIKEN, 
     Wako 2-1, Saitama 351-0198, Japan  
\\
$^3$ Institute of Physics, University of Tsukuba, 
     Tsukuba, Ibaraki 305-8571, Japan 
}

\vspace{5mm}

(Received May 10, 2004)
  
\vspace{10mm}   
\end{center}   
  
\begin{center}   
\begin{minipage}{12cm}   
\baselineskip 16pt   
\noindent   
 The nonperturbative aspects of string theory are explored 
for non-critical string in two distinct formulations, 
loop equations and matrix models. 
The effects corresponding to the D-brane in these formulations 
are especially investigated in detail. 
 It is shown that matrix models can universally yield 
a definite value of the chemical potential for an instanton 
while loop equations cannot. 
 This implies that it may not be possible to formulate 
string theory nonperturbatively 
solely in terms of closed strings.

\end{minipage}   
\end{center}   
%
\newpage   
\baselineskip 18pt   
\setcounter{footnote}{0}  
\renewcommand{\theequation}{\thesection.\arabic{equation}}
\renewcommand{\thefootnote}{\arabic{footnote}}
\section{Introduction} 
\label{sec:intro} 
\setcounter{equation}{0}
 It is now widely recognized that 
the nonperturbative effect in string theory that behaves as  
$\sim e^{-\frac{1}{g_s}}$ \cite{Shenker:1990uf} 
stems from the dynamical hypersurfaces 
in space-time (the D-brane) 
on which open strings can end \cite{Polchinski:fq}. 
The double scaling limit of matrix models 
\cite{Brezin:rb,MMreview} 
may enable us to learn more about this, at least for noncritical strings. 
 It was realized in the early 1990s  
that the string equation 
(Painlev\'e equation) indeed inherits 
such nonperturbative effects \cite{David:sk,Eynard:1992sg}; 
they correspond to the deviation of its solutions 
from the genus expansion 
of two-dimensional gravity. 
 In the early 2000s, 
studies of the D-brane have been developed 
from the viewpoint of Liouville field theory 
\cite{Fateev:2000ik,Zamolodchikov:2001ah,Ponsot:2001ng}. 

 Given this situation, 
we are tempted to reinvestigate 
nonperturbative effects of the noncritical string theory 
in full detail. 
 In particular, we would like to ask 
if nonperturbative effects introduce a continuous parameter 
characterizing vacua like the $\theta$ parameter in QCD, 
or if they are fully calculable as an intrinsic property 
of the matrix model.
 
 We focus on the $c=0$ noncritical string here 
in order to address the above questions as clearly as possible. 
 First, let us recall the nonperturbative effects 
observed in string equation, 
following Refs.~\cite{David:sk} and \cite{Eynard:1992sg}. 
 The variable for which to solve is the specific heat $u(t)$, 
depending on the renormalized cosmological constant $t$. 
 The string equation with respect to it 
takes the form of Painlev\'e equation I, 
\begin{equation} 
 u^2 + \frac{g_s^2}{6}\,\del_t^2 u = t \, . 
  \label{eq:string_equation}
\end{equation} 
 The leading-order contribution to $u(t)$ in the genus expansion 
is $(g_s)^0$, and it is related to the free energy $F$ 
($F \equiv {\rm ln}\,Z$ 
with the matrix model partition function $Z$) through 
\begin{equation} 
 u = g_s^2\, \del_t^2 F \, . 
\end{equation} 
$u(t)$ admits a perturbative series expansion: 
\begin{equation} 
 u_{\rm pert}(t) 
 = 
 - \sqrt{t} 
 + \frac{1}{48}\,\frac{g_s^2}{t^2} 
 + \frac{49}{4608}\,\frac{g_s^4}{t^{\frac{9}{2}}} 
 + \cdots \, . 
\end{equation} 
 However, there may be some degree of ambiguity in the above solution, 
due to nonperturbative effects. 
 To examine this point, 
let us suppose that there is another solution $u(t)$ 
infinitesimally close to the above perturbative solution, which we write 
\begin{equation} 
 u(t) = u_{\rm pert}(t) + \Delta u(t) \, .
  \label{eq:def_of_delta_u}
\end{equation} 
Then, if we write $\Delta u(t)$ in the form 
\begin{equation} 
 \Delta u(t) = A\, e^{-\frac{h(t)}{g_s}} \, , 
  \label{eq:def_of_h} 
\end{equation} 
the equation for $h(t)$ admits an expansion with respect to $g_s$, 
\begin{equation} 
 \left((\del_t h(t))^2 - 12 \sqrt{t} +A e^{-\frac{h(t)}{g_s}}\right) 
 + 
 g_s \left(- \del_t^2 h(t) \right) 
 + {\cal O}(g_s^2) 
 = 0 \, . 
\end{equation} 
 Noting that  $e^{-\frac{h(t)}{g_s}}$ is negligible 
as long as $g_s$ is small,  $h(t)$ can be solved as 
\begin{equation} 
 h(t) 
 = 
 \frac{8\sqrt{3}}{5}\,t^{\frac{5}{4}} 
 + 
 \frac{g_s}{8}\,{\rm ln}\,t 
 + 
 {\cal O}(g_s^2) \, . 
\end{equation} 
 By inserting this into 
Eqs.~(\ref{eq:def_of_h}) and (\ref{eq:def_of_delta_u}) 
and integrating $u(t)$ twice over $t$, 
the free energy $F$ is found as 
\begin{eqnarray} 
 && 
 F = F_{\rm pert} + F_{\rm inst} \, , 
  \nonumber \\ 
 && 
 F_{\rm pert} 
 = 
 - \frac{4}{15}\, \frac{t^{\frac{5}{2}}}{g_s^2} 
 - \frac{1}{48}\,{\rm ln}\,t 
 + \frac{7}{5760}\,\frac{g_s^2}{t^{\frac{5}{2}}} 
 + {\cal O}(g_s^4) \, , 
  \nonumber \\ 
 && 
 F_{\rm inst} 
 = 
 \frac{C}{t^{\frac{5}{8}}}\, 
 \exp 
 \left[ 
  - \frac{8 \sqrt{3}}{5 g_s}\,t^{\frac{5}{4}} 
 \right] 
 \left( 
  1 + {\cal O}(g_s) 
 \right) \, .  \label{eq:finst}
\end{eqnarray} 
 Here, $F_{\rm inst}$ is the non-perturbative contribution 
to the free energy, which can be interpreted as an instanton effect
\cite{David:sk,Polchinski:fq}. 
 Recent developments have identified the origin of these effects 
as D-branes 
\cite{McGreevy:2003kb,Martinec:2003ka,McGreevy:2003ep,
Klebanov:2003km,Alexandrov:2003nn}. 
 However, 
the overall factor of $C$ 
for the instanton contribution cannot be determined 
from the string equation (\ref{eq:string_equation}) itself; 
it is a constant of integration that should be determined 
from the other boundary condition mentioned above. 
That is, the string equation determines 
the D-instanton action 
\begin{equation} 
 S_{\rm inst} 
 = 
 \frac{8 \sqrt{3}}{5g_s}t^{\frac54}
 \label{eq:sinstpv}
\end{equation} 
uniquely, 
and it shows that the D-instanton contributes to the free energy 
in the form 
\begin{equation} 
 \frac{C}{t^{\frac{5}{8}}}\,e^{-S_{\rm inst}} \, .  
 \label{eq:Dinst}
\end{equation} 
However, 
the weight $C$ relative to the perturbative contribution, 
which corresponds to the chemical potential 
needed to bring a D-instanton into the system, 
remains undetermined with the string equation 
(\ref{eq:string_equation}). 

 Our main focus here is concerned with the chemical potential 
of the D-instanton. 
 There are the following two possibilities for $C$: 
\begin{description} 
 \item[(a)] 
  $C$ is a parameter characterizing the vacua 
as a result of quantizing the system, like the $\theta$-parameter in QCD. 
 Specifically, each value of $C$ 
corresponds to a distinct vacuum. 
 \item[(b)] 
  $C$ is calculable and its value is uniquely determined. 
  In this sense, 
the string equation does not 
fully describe the properties of the system. 
\end{description} 

 We are able to determine which possibility is actually realized 
by inquiring whether the chemical potential of the instanton 
can be calculated directly 
by carrying out the path integral of the $c=0$ matrix model. 
 
 Our main result here is the following: 
The value of $C$ can be calculated. 
  Moreover, 
$C$ does not depend on the details of the matrix model action, 
and it is thus a {\it universal} quantity, given by 
\begin{equation} 
 C = i\,\frac{1}{8 \cdot 3^{\frac{3}{4}} \cdot \sqrt{\pi}} \, . 
  \label{eq:value_of_chemical_pot}
\end{equation} 
 Reflecting the instability of the vacuum 
in the presence of a D-instanton, $C$ is a purely imaginary number. 
 Our result shows that the lifetime of such a vacuum 
is uniquely determined and does not depend on the regularization scheme. 
 These results indicate that assertion {\bf (b)} is correct. 

 This paper is organized as follows. 
 We start in \S\ref{sec:inst} 
by identifying the instanton effect in the matrix model 
and showing that it corresponds to the D-instanton effect 
in $c=0$ Liouville field theory. 
Then, we 
 divide the partition function of the matrix model 
into the sum of contributions $Z^{(n\text{-inst})}$ 
from $n$ of instantons 
($n = 0,\,1,\,\cdots$) 
by restricting the integration regions of the eigenvalues 
in a proper way. 
 There, we also derive various significant equations 
for estimating the chemical potential $\mu$ of an instanton, 
i.e., 
the ratio of the single-instanton contribution $Z^{(1\text{-inst})}$ 
to the trivial vacuum contribution $Z^{(0\text{-inst})}$, 
\begin{equation} 
 \mu \equiv 
 \frac{Z^{(1\text{-inst})}}{Z^{(0\text{-inst})}} \, . 
\end{equation} 
 In \S\ref{sec:chemical_pot}, 
we apply the expression obtained in \S\ref{sec:inst} 
to the estimation of the chemical potential of an instanton, 
with the help of the orthogonal polynomial method. 
 We find that the chemical potential of an instanton is indeed given by 
the value in Eq.~(\ref{eq:value_of_chemical_pot}). 

 Section \ref{sec:loop_eq} constitutes the second part of the paper. 
Then, we reconsider the result obtained in the previous sections 
from the matrix model calculation 
by examining the loop equations more closely. 
 First, we consider the loop amplitude 
in the background produced by single-instanton. 
 This is done by taking the expectation value of the loop amplitude 
after restricting ourselves to single-instanton sector a priori, 
ignoring the other sectors. 
 We find that 
the loop equations in the large $N$ limit 
determine the loop amplitude in the single-instanton sector, 
up to a constant factor. 
 In order to determine this constant, 
we need the loop equations at {\it all} orders. 

 Next, we examine 
whether the loop equations determine 
the chemical potential of an instanton. 
 As seen in \S\ref{sec:chemical_pot}, 
the chemical potential of an instanton is universal. 
 However, it requires regularization of a divergence of type 
${\rm ln}\,N$. 
 Thus, we find that the loop equations obtained after taking 
the continuum limit 
cannot determine the chemical potential of an instanton. 
 This fact indicates  that it is quite difficult in 
string theory with closed strings only 
to fully describe 
the nonperturbative effects of strings, 
and that open strings or matrices 
should be regarded as fundamental degrees of freedom. 
 Section \ref{sec:conclusion} contains the conclusion of this paper. 

\section{\boldmath The instanton in $c=0$ noncritical string theory}
\label{sec:inst} 
\setcounter{equation}{0}
In this section, we show that the instanton in the one-matrix model 
is identical to the D-instanton in the $c=0$ noncritical string theory 
and that it indeed gives the nonperturbative effect discussed 
in the introduction.

\subsection{Action of the instanton}
\label{subsec:Sinst}

\begin{figure}[htbp]
\begin{center}
\includegraphics{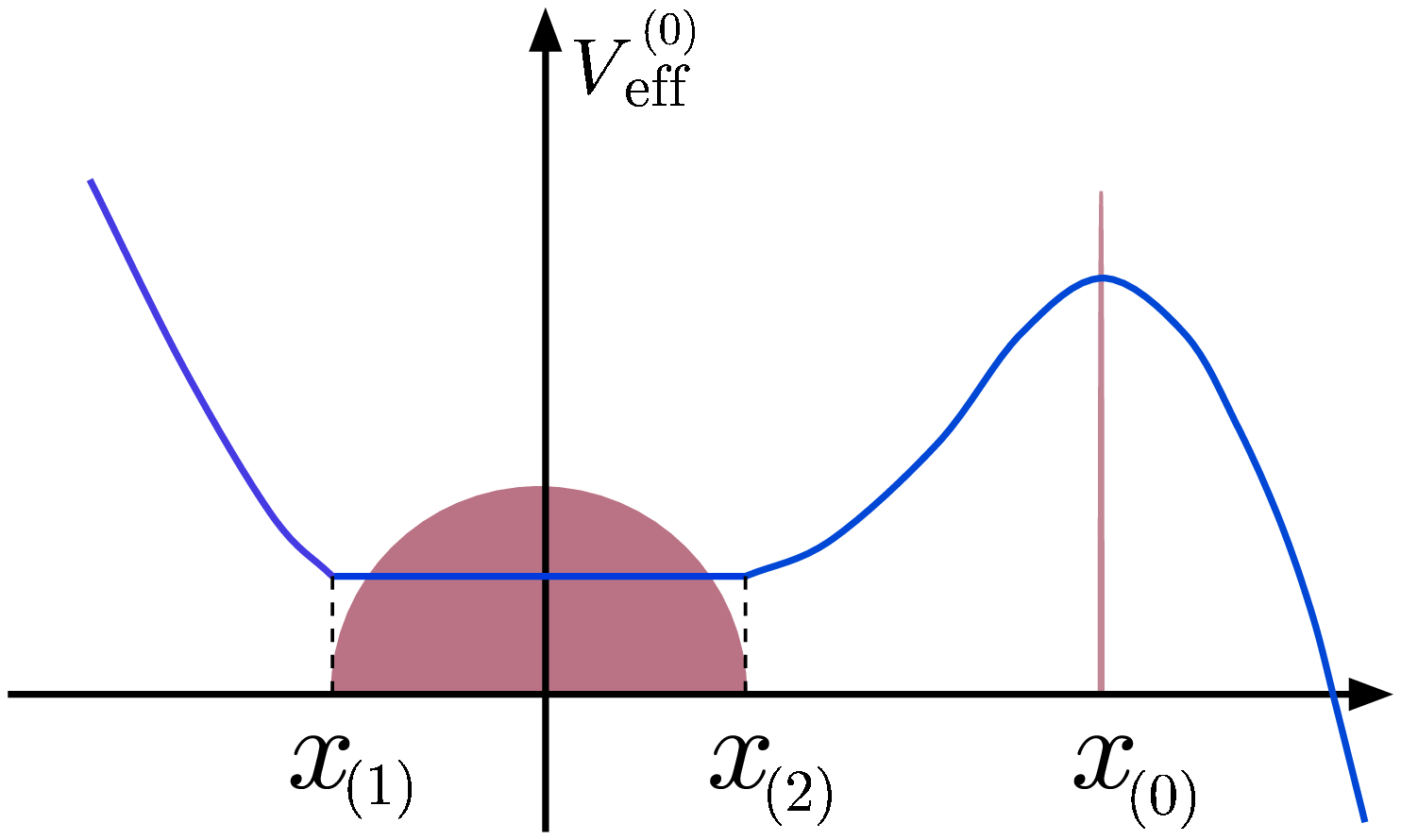}
\caption{$\Veff^{(0)}$.}
\label{fig:veffba}
\end{center}
\end{figure}

As a concrete example, we consider the one-matrix model 
with a cubic potential,
\begin{equation}
S=N\tr V(\phi), \hspace{10mm}V(x)=\frac{1}{2}\phi^2-\frac{g}{3}\phi^3,
\label{eqn:phi3}
\end{equation}
where $\phi$ is an $N\times N$ Hermitian matrix. 
The effective action for the eigenvalues $\lambda_i$ ($i=1,\cdots,N$) 
is given by 
\begin{equation}
\Seff=-\sum_{i<j}\log (\lambda_i-\lambda_j)^2
     +N\sum_{i} V(\lambda_i).
\label{eqn:Seff}              
\end{equation}
Hereafter, we consider the situation 
in which a single eigenvalue ($\lambda_{N}$) is separated from the others. 
Then, the partition function of the matrix model (\ref{eqn:phi3}) 
is expressed as 
\begin{eqnarray}
Z_N & = & \int dx \int d\lambda_1\cdots d\lambda_{N-1}
          \left(\prod _{i=1}^{N-1} (x-\lambda_i)^2 \right)
          \Delta^{(N-1)}(\lambda_1,\cdots,\lambda_{N-1})^2 \nonumber \\
    &   & \times e^{-N\sum_{i=1}^{N-1}V(\lambda_i)}e^{-NV(x)}, 
\label{eqn:Z}    
\end{eqnarray}
where $x$ denotes $\lambda_N$, and 
$\Delta^{(N-1)}(\lambda_1,\cdots,\lambda_{N-1})^2$ is the 
Vandermonde determinant in terms of $\lambda_1,\cdots,\lambda_{N-1}$.
Thus we have $\left(\prod _{i=1}^{N-1} (x-\lambda_i)^2 \right)
\Delta^{(N-1)}(\lambda_1,\cdots,\lambda_{N-1})^2=\Delta^{(N)}(\lambda_1,\cdots,\lambda_{N})^2$. 
By introducing an $(N-1)\times (N-1)$ Hermitian matrix $\phi'$, 
this can be rewritten as 
\begin{eqnarray}
Z_N & = & Z'_{N-1}\int dx \left\langle \det(x-\phi')^2\right\rangle'
          e^{-NV(x)} \nonumber \\
     &\equiv & Z'_{N-1}\int dx e^{-N\Veff(x)},
\end{eqnarray}
where
\begin{eqnarray}
Z'_{N-1} & = & \int d\phi' e^{-N\tr V(\phi')}, 
\nonumber \\
\left\langle O \right\rangle'
 & = & \frac{1}{Z'_{N-1}}\int d\phi' O e^{-N\tr V(\phi')}.
\label{eqn:(N-1)mm} 
\end{eqnarray}
In the large-$N$ limit, we can set 
\begin{equation}
\left\langle\frac{1}{N-1}\tr\left(\frac{1}{x-\phi'}\right)\right\rangle'=R(x),
\label{eqn:(N-1)resolvent}
\end{equation}
where $R(x)$ is the resolvent of the matrix model (\ref{eqn:phi3}), 
which is given by 
\begin{eqnarray}
R(x) & = & \left\langle\frac{1}{N}\tr\frac{1}{x-\phi}\right\rangle 
\nonumber \\
 & = & \frac{1}{2}\left(V'(x)+\sqrt{V'(x)^2+p(x)}\right). 
\label{eqn:resolvent}
\end{eqnarray}
Here, $p(x)$ is a polynomial of degree 1, and the branch 
of the square root is taken so that $R(x)\sim 1/x$ as 
$x\rightarrow\infty$. 
Therefore, in the large-$N$ limit, $\Veff$ becomes 
\begin{eqnarray}
\Veff^{(0)}(x) & = & -2 \Repa \int_{x_*}^{x}dx' R(x') +V(x) 
\nonumber \\
               & = & -\Repa \int_{x_*}^{x}dx' \sqrt{V'(x')^2+p(x')},
\label{eqn:Veff0}            
\end{eqnarray}
where $x_*$ fixes the origin of $\Veff^{(0)}$.  
Here, we note that when $x'$ in the above integration is 
inside the region where the other eigenvalues are distributed, 
we have to take the real part of the resolvent, 
because the square of the characteristic polynomial 
$\det(x'-\phi')^2$ is originally positive-definite.  
Accordingly, $\Veff^{(0)}(x)$ has a plateau in this region, 
as shown in Fig.~\ref{fig:veffba}. 
The meaning if this plateau is that the force from the other eigenvalues 
acting on the eigenvalue we are considering vanishes 
in this region. 
This is natural, because inside the sea of eigenvalues, the force 
should be balanced to zero. 
Thus if we choose a value of $x$, $x_*$, inside the plateau, 
$\Veff^{(0)}(x)$ gives the effective potential 
in the case that we regard its origin as the value at the plateau. 
In particular, its local maximum $\Veff^{(0)}(x_{(0)})$ can be regarded 
as the potential of the instanton that lies at the top of the 
effective potential $\Veff^{(0)}$. A simple computation shows 
that in the double scaling limit\footnote
{Here we have set the string coupling constant as $g_s^2=1$. 
thus, all dimensional quantities in the continuum limit 
become dimensionless in units of $g_s$. In \S\ref{sec:inst} 
and \S\ref{sec:chemical_pot}, we follow this convention.} we have 
\begin{equation}
g=g_\star\exp(-\epsilon^2 t),~~~x=x_\star\exp(\epsilon\zeta),~~~
N^2=\epsilon^{-5}.
\label{eqn:dsl}
\end{equation}
The height of the potential barrier then becomes 
\begin{equation}
N(\Veff^{(0)}(x_{(0)})-\Veff^{(0)}(x_{(2)}))=\frac{8\sqrt{3}}{5}t^{\frac{5}{4}},
\label{eqn:Sinst}
\end{equation}
where $t$ is normalized so that the `specific heat' $u$ 
derived from the free energy $F(t)$ as $u=\ddot{F}$ satisfies 
the Painlev$\acute{\mbox{e}}$ equation, 
\begin{equation}
u^2+\frac{1}{6}\ddot{u}=t.
\end{equation}
Correspondingly, $\zeta$ is normalized so that the disk amplitude 
obtained by taking the scaling limit of (\ref{eqn:resolvent}) 
becomes 
\begin{equation}
\tilde{w}(\zeta)=\left(\zeta-\frac{1}{2}\sqrt{t}\right)
           \sqrt{\zeta+\sqrt{t}}.
\label{eqn:diskamp}           
\end{equation}
Note that (\ref{eqn:Sinst}) agrees with the prediction 
of the Painlev$\acute{\mbox{e}}$ equation, {eq:sinstpv}) 
\cite{David:sk}. 
In \S\ref{sec:chemical_pot}, we examine how the coefficient 
as well as the power of $t$ in (\ref{eq:finst}) can be 
determined from the point of view of the matrix model.

\subsection{Chemical potential of the instanton}

Next we take a closer look at the contribution from the instanton 
to the free energy. 
We begin with the partition function $Z_N$, 
which can be written as 
\begin{equation}
Z_N=\int d\lambda_1\cdots d\lambda_N G,~~~
G=\Delta^{(N)}(\lambda_1,\cdots,\lambda_N)^2
  \exp\left(-N\sum_{i=1}^{N}V(\lambda_i)\right).
\end{equation}
This can be divided 
into the sum of the multi-instanton sectors 
in the large-$N$ limit, where the 
interactions between the instantons are of ${\cal O}(1/N)$ 
compared to the leading contributions, as we see below.
Thus we write 
\begin{equation}
Z_N=Z_N^{(0\text{-inst})}+Z_N^{(1\text{-inst})}+Z_N^{(2\text{-inst})}+\cdots, 
\end{equation}
where the $k$-instanton sector is characterized as that 
in which $k$ eigenvalues are separated from others; that is, 
they do not lie in the region $x_{(1)}<x<x_{(2)}$. 

 Let us consider the 1-instanton sector:
\begin{eqnarray}
Z_N^{(1\text{-inst})}
 & = & N\int_{x<x_{(1)}, x_{(2)}<x} dx 
        \int_{x_{(1)} \le \lambda_i \le x_{(2)}~(i\neq N)}
        \prod_{i=1}^{N-1}d\lambda_i 
        G(x,\lambda_1,\cdots, \lambda_{N-1})  
\nonumber \\
 & = & N{Z'_{N-1}}^{(0\text{-inst})}\int_{x<x_{(1)}, x_{(2)}<x} dx 
       \left\langle \det (x-\phi')^2 
       \right\rangle^{\prime \,(0\text{-inst})}
       e^{-NV(x)} 
\nonumber \\
 & = & N{Z'_{N-1}}^{(0\text{-inst})}\int_{x<x_{(1)}, x_{(2)}<x} dx f(x), 
\label{eqn:1-inst.F} 
\end{eqnarray}
where the overall factor $N$ reflects the number of ways 
of specifying the isolated eigenvalue. Quantities 
with primes are defined as in (\ref{eqn:(N-1)mm}):
\begin{eqnarray}
{Z'_{N-1}}^{(0\text{-inst})} & = & \int d\phi' e^{-N\tr V(\phi')}, 
\nonumber \\
\left\langle O \right\rangle^{\prime \, (0\text{-inst})}
 & = & \frac{1}{{Z'_{N-1}}^{(0\text{-inst})}}
 \int d\phi' O e^{-N\tr V(\phi')}. \label{eq:average_0-inst}
\end{eqnarray}
Here, all eigenvalues of $\phi'$ are understood to lie 
between $x_{(1)}$ and $x_{(2)}$, and 
\begin{equation}
f(x)= \left\langle \det (x-\phi')^2 
      \right\rangle^{\prime \, (0\text{-inst})}
       e^{-NV(x)}.
\end{equation}
It is evident that in (\ref{eqn:1-inst.F}) if we change the interval 
of the integration with respect to $x$ from $\{x<x_{(1)}, x_{(2)}<x\}$ 
to $\{x_{(1)}<x<x_{(2)}\}$, this integral would give 
the 0-instanton partition function 
multiplied by $N$. This observation leads us to the relation 
\begin{equation}
N{Z'_{N-1}}^{(0\text{-inst})}\int_{x_{(1)}<x<x_{(2)}} dx f(x)
=NZ_N^{(0\text{-inst})}.
\end{equation}
Thus, we derive the chemical potential of the instanton 
in terms of the correlator of the matrix model (\ref{eqn:phi3}) as 
\begin{equation}
\mu\equiv\frac{Z_N^{(1\text{-inst})}}{Z_N^{(0\text{-inst})}}
=N\frac{\int_{x<x_{(1)}, x_{(2)}<x} dx \left\langle \det (x-\phi)^2 \right\rangle^{(0\text{-inst})}e^{-NV(x)}}
{\int_{x_{(1)}<x<x_{(2)}} dx \left\langle \det (x-\phi)^2 \right\rangle^{(0\text{-inst})}e^{-NV(x)}},
\label{eqn:chempot}
\end{equation}
where we have used the fact that in the large-$N$ limit (or the double
scaling limit) we have 
\begin{equation}
f(x)
=\left\langle \det (x-\phi)^2 \right\rangle^{(0\text{-inst})}
 e^{-NV(x)},
\end{equation}
as in (\ref{eqn:(N-1)resolvent}). 

Similarly, $Z_N^{(k\text{-inst})}$ is given as 
\begin{eqnarray}
Z_N^{(k\text{-inst})}
 & = & {}_NC_k\left({Z'_{N-k}}^{(0\text{-inst})}\right)^k 
 \nonumber \\
 & \times &   \prod_{i=N-k+1}^{N}
       \left(\int_{x_i<x_{(1)}, x_{(2)}<x_i} dx_i
       \left\langle \det(x_i-\phi')^2 
       \right\rangle^{\prime \, (0\text{-inst})} 
       e^{-NV(x_i)} \right) \nonumber \\
 & & \times \Delta^{(k)}(x_{N-k+1},\cdots,x_{N})^2, 
\end{eqnarray}
where the prime indicates quantities of the $(N-k)\times (N-k)$ 
matrix model. As long as $k\ll N$, the last factor, 
$\Delta^{(k)}(x_{N-k+1},\cdots,x_{N})^2$, can be ignored, 
because it gives only an ${\cal O}(1/N)$ contribution compared to 
the leading one. Physically, this corresponds to switching off 
interactions between instantons (the dilute gas approximation).
Then, the above equation becomes 
\begin{equation}
Z_N^{(k\text{-inst})}
={}_NC_k\left({Z'_{N-k}}^{(0\text{-inst})}\right)^k
 \left(\int_{x<x_{(1)}, x_{(2)}<x} dx f(x)\right)^{k}.
\end{equation}
Repeating the same argument, we obtain 
\begin{equation}
{}_NC_k\left({Z'_{N-k}}^{(0\text{-inst})}\right)^k
\left(\int_{x_{(1)}<x<x_{(2)}} dx f(x)\right)^{k}
={}_NC_kZ_N^{(0\text{-inst})}
\end{equation}
and 
\begin{equation}
\frac{Z_N^{(k\text{-inst})}}{Z_N^{(0\text{-inst})}}
={}_NC_k\left(\frac{\int_{x<x_{(1)}, x_{(2)}<x} dx \left\langle \det (x-\phi)^2 \right\rangle^{(0\text{-inst})}e^{-NV(x)}}
{\int_{x_{(1)}<x<x_{(2)}} dx \left\langle \det (x-\phi)^2 \right\rangle^{(0\text{-inst})}e^{-NV(x)}}\right)^{k}.
\end{equation}
This implies that the free energy is in fact given by 
\begin{equation}
F=F^{(0\text{-inst})}
 +\mu.
\end{equation}

\subsection{An instanton as a D-instanton}

In this subsection we confirm that the instanton of the matrix model 
we have considered is indeed the D-instanton in the $c=0$ noncritical string theory. 

We first note that in ordinary critical string theory, 
contributions from a D-brane correspond 
to adding surfaces with open boundaries. Let us check that 
this is also the case with our instanton. 
We rewrite the partition function (\ref{eqn:Z}) 
as 
\begin{equation}
Z_N=\int dx\int d\phi' dq d\bar{q}~e^{-S^{(1-{\rm inst.)}}-NV(x)},
\end{equation}
where 
\begin{equation}
S^{(1\text{-inst})}=N\tr V(\phi')+\sum_{i=1,2}\bar{q}_i(\phi'-x)q_i,
\label{eqn:qqbar}
\end{equation}
and $q_i$ is Grassmann odd in the fundamental representation 
of $U(N-1)$. 
This implies that if we evaluate the integration over $x$ 
by using the saddle point value at $x=x_{(0)}$, 
the interactions over $q$ and $\bar{q}$ 
provide new contributions from surfaces with open boundaries \cite{Yang:hs,Minahan:1992bz,Kazakov:1991pt}
for the partition function. This can be expressed as  
\begin{equation}
\Delta Z_N^{(1\text{-inst})}
=\raisebox{- 35pt}{\scalebox{0.4}{\includegraphics{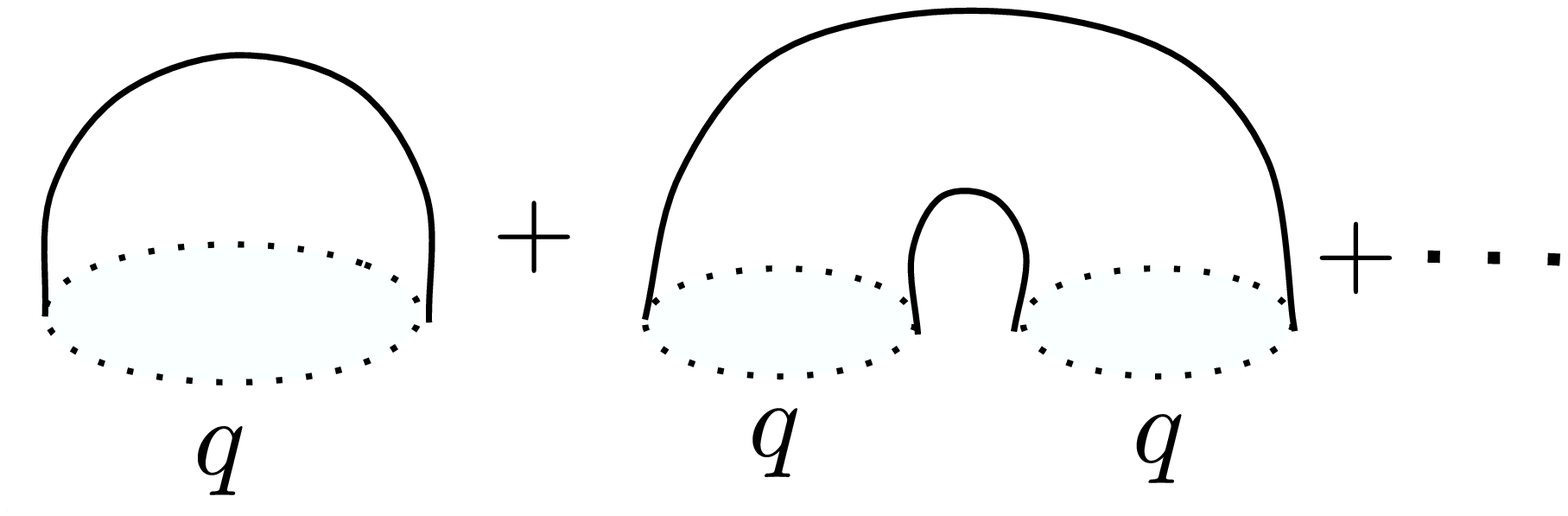}}}
 \,\, .
\label{eqn:qeffect} 
\end{equation}
From these considerations, we are led to suspect 
that our instanton is identically the D-instanton.

In order to make a precise comparison with the continuum (Liouville) 
theory, let us consider a loop amplitude in the instanton background. 
If we take account of the effect of the instanton, the resolvent 
can be written 
\begin{eqnarray}
\lefteqn{\left\langle\frac{1}{N}\tr\left(\frac{1}{z-\phi}\right)
         \right\rangle} \nonumber \\
 & = & \frac{1}{Z_N}\int dx\int d\lambda_1\cdots\lambda_{N-1} 
 \frac{1}{N}\left(\frac{1}{z-x}
                  +\sum_{i=1}^{N-1}\frac{1}{z-\lambda_i}\right)
 \nonumber \\
 & & \times \prod_{i=1}^{N-1}(x-\lambda_i)^2
     \Delta^{(N-1)}(\lambda_1,\cdots,\lambda_{N-1})^2
     e^{-N\sum_{i=1}^{N-1}V(\lambda_i)}e^{-NV(x)} \nonumber \\
 & \sim & \frac{Z_N^{(0\text{-inst})}}{Z_N}
          \left\langle \frac{1}{N}\tr\left(\frac{1}{z-\phi}\right)
          \right\rangle^{(0\text{-inst})}
         +\frac{Z_N^{(1\text{-inst})}}{Z_N}
          \left(\frac{1}{N}\frac{1}{z-x_{(0)}}
               +\left\langle 
                \frac{1}{N}\sum_{i=1}^{N-1}\frac{1}{z-\lambda_i}    
                \right\rangle_{x_{(0)}}\right) \nonumber \\
 & &     +\cdots,                 
\label{eqn:totalresolvent}
\end{eqnarray}
where we have replaced the integration 
with respect to $x$ over the interval $x<x_{(1)}$, 
$x_{(2)}<x$ with the saddle point value at $x=x_{(0)}$, which is valid 
in the large-$N$ limit, and we have used 
\begin{eqnarray}
\left\langle\frac{1}{N}\sum_{i=1}^{N-1}\frac{1}{z-\lambda_i}
\right\rangle_{x_{(0)}} 
 & = & 
 \frac{1}{Z_N^{(1\text{-inst})}}N\int d\lambda_1\cdots\lambda_{N-1}
 \left(\frac{1}{N}\sum_{i=1}^{N-1}\frac{1}{z-\lambda_i}\right) 
 \nonumber \\
 & \times & \prod_{i=1}^{N-1}(x_{(0)}-\lambda_i)^2
            \Delta^{(N-1)}(\lambda_1,\cdots,\lambda_{N-1})^2
            e^{-N\sum_{i=1}^{N-1}V(\lambda_i)}e^{-NV(x_{(0)})}. 
 \nonumber \\           
\label{eqn:backreaction}     
\end{eqnarray} 
{}From (\ref{eqn:totalresolvent}), the resolvent 
in the 1-instanton background is given by 
\begin{equation}
\left\langle \frac{1}{N}\tr\left(\frac{1}{z-\phi}\right)
          \right\rangle^{(1\text{-inst})}
  =   \frac{1}{N}\frac{1}{z-x_{(0)}}
       +\left\langle\frac{1}{N}\sum_{i=1}^{N-1}\frac{1}{z-\lambda_i}    
        \right\rangle_{x_{(0)}}.
\label{eqn:1-instresolvent}
\end{equation}
As discussed below [see (\ref{eqn:Rexp})], this gives the resolvent 
in the 0-instanton sector at leading-order 
in the large-$N$ limit as 
\begin{equation}
\left\langle \frac{1}{N}\tr\left(\frac{1}{z-\phi}\right)
\right\rangle^{(1\text{-inst})} 
=\left\langle \frac{1}{N}\tr\left(\frac{1}{z-\phi}\right)
 \right\rangle^{(0\text{-inst})}
+\Delta R^{(1\text{-inst})}, 
\label{eqn:0-instRin1-instR}
\end{equation}
\begin{equation}
\Delta R^{(1\text{-inst})}
\equiv\frac{1}{N}\frac{1}{z-x_{(0)}}
+\left(\left\langle
            \frac{1}{N}\sum_{i=1}^{N-1}\frac{1}{z-\lambda_i}    
       \right\rangle_{x_{(0)}}
-\left\langle \frac{1}{N}\tr\left(\frac{1}{z-\phi}\right)
       \right\rangle^{(0\text{-inst})}
       \right).
\label{eqn:DeltaR}
\end{equation}
In $\Delta R^{(1\text{-inst})}$, we see that 
the first term, $1/N\cdot1/(z-x_{(0)})$, corresponds to 
the isolated eigenvalue, which is distributed at $x=x_{(0)}$ 
as a $\delta$-function, 
and the second term indicates that 
this eigenvalue causes a distortion of the distribution 
of the other eigenvalues, as shown by the factor 
$\prod_{i=1}^{N-1}(x_{(0)}-\lambda_i)^2$ in (\ref{eqn:backreaction}), 
both of which turn out to be at next-to-leading order. 
The physical meaning of $\Delta R^{(1\text{-inst})}$ is quite clear. 
Because it is precisely the contribution from the instanton 
to the loop amplitude, it can also be computed by using the action 
(\ref{eqn:qqbar}) as surfaces with loops of $q$ and $\bar{q}$.  
Thus, it is found that $\Delta R^{(1\text{-inst})}$ 
describes the correction to the loop amplitude from open boundaries: 
\begin{equation}
\Delta R^{(1\text{-inst})}
=\raisebox{- 35pt}{\scalebox{0.4}{\includegraphics{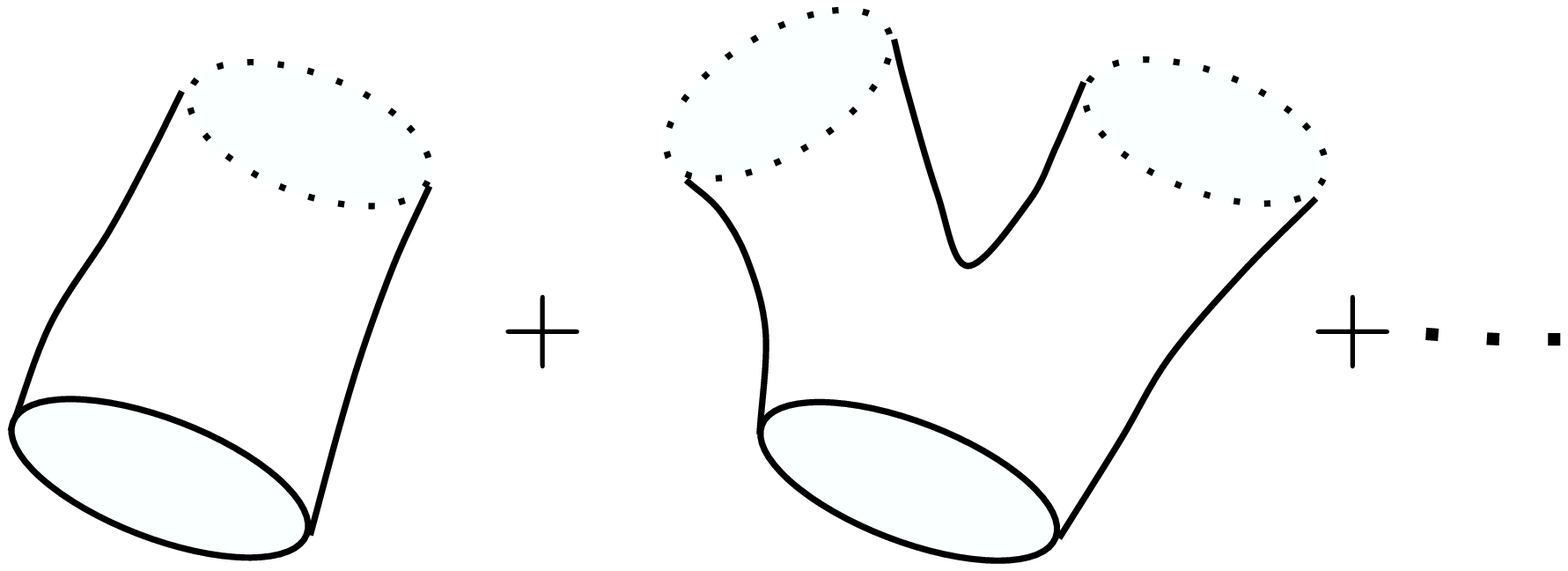}}} 
,
\end{equation}

It is not difficult to compute $\Delta R^{(1\text{-inst})}$ 
explicitly. 
First, we note that the resolvent is also given 
by (\ref{eqn:resolvent}), even in the instanton background, 
because it can be derived generically 
using the Schwinger-Dyson equation. 
Thus, a different choice of the polynomial $p(z)$ 
yields a description of a different instanton sector. 
Because the behavior of $R(z)$ 
as $z \rightarrow \infty$ fixes the coefficient of the first-order term 
in $p(z)$, it is a constant term that distinguishes the instanton
background. 
Denoting $p(z)$ in the absence of the instanton 
by $p_0(z)$ , the resolvent in the instanton background is thus 
given by  
\begin{eqnarray}
R(z) & = & \frac{1}{2}\left(\sqrt{V'(z)^2+p_0(z)+c}+V'(z)\right) 
\nonumber \\
 & = & R_0(z)+\frac{c}{4}\frac{1}{\sqrt{V'(z)^2+p_0(z)}}+\cdots, 
\label{eqn:Rexp} 
\end{eqnarray}
where $R_0(z)$ is the resolvent in the absence of the instanton, 
and $c$ is a certain constant of ${\cal O}(1/N)$ determined below. 
The above expansion corresponds to the right-hand side 
of (\ref{eqn:0-instRin1-instR}) term by term. 
Therefore, $\Delta R^{(1\text{-inst})}(z)$ takes the form 
\begin{equation}
\Delta R^{(1\text{-inst})}(z)
=\frac{c}{4}\frac{1}{\sqrt{V'(z)^2+p_0(z)}}+\cdots .
\label{eqn:1-instloop}
\end{equation}
In order to find the value of $c$ 
in the case of the single-instanton background, 
we note that the eigenvalue density $\rho(x)$ is related to $R(z)$ 
as
\begin{equation}
\rho(x)=-\frac{1}{\pi}\Impa R(x+i0). \label{eq:rhodef}
\end{equation}
Using this, we impose the condition that the integration of $\Delta \rho(x)$ 
corresponding to $\Delta R^{(1\text{-inst})}(z)$ 
on the small interval around $z=x_{(0)}$ yields $1/N$, 
which implies that a single eigenvalue is located at $z=x_{(0)}$. 
This condition amounts to requiring 
\begin{equation}
\frac{1}{2\pi i}\oint_{C_{x_{(0)}}} dz \Delta R^{(1\text{-inst})}(z)
=\frac{1}{N},    
\end{equation}
where $C_{x_{(0)}}$ is a small circle surrounding $z=x_{(0)}$. 
This gives 
\begin{equation}
c=\frac{8}{N}W'_0(x_{(0)}),
\end{equation}
where 
\begin{equation}
W_0(z)=\frac{1}{2}\sqrt{V'(z)^2+p_0(z)}.
\end{equation}
Substituting this into (\ref{eqn:1-instloop}), we find
\begin{equation}
\Delta R^{(1\text{-inst})}(z)
=\frac{1}{N}\frac{W'_0(x_{(0)})}{W_0(z)}+\cdots .
\end{equation}
In the double scaling limit, this quantity remains finite, 
and the first term gives 
\begin{equation}
\frac{3\sqrt{3}}{8}\frac{t^{\frac{1}{4}}}{\tilde{w}(\zeta)}
\label{eqn:DeltaRvalue}
\end{equation}
as the instanton correction to the disk amplitude 
in the absence of the instanton $\tilde{w}(\zeta)$ given 
in (\ref{eqn:diskamp}). It is easy to check that we can obtain 
exactly the same result obtained from the case of the $\phi^4$ potential. 
This implies that (\ref{eqn:DeltaRvalue}) is universal. 
In \S\ref{sec:loop_eq}, we see that (\ref{eqn:DeltaRvalue}) 
coincides with the results in the Liouville theory, or 
the loop equations. Therefore, we find that the instanton 
is identical to the D-instanton in $c=0$ noncritical string theory.

\section{Chemical potential of the instanton} 
\label{sec:chemical_pot} 
\setcounter{equation}{0}
We have seen that we can compute 
various quantities in the D-instanton background 
by using the matrix model in the previous section.
In this section, we show further that the matrix model 
also makes it possible to obtain a definite value 
of the chemical potential for the instanton (\ref{eqn:chempot}), 
namely, the weight of the instanton itself. 

 We start by computing the numerator 
in Eq.~(\ref{eqn:chempot}). 
 For $x < x_{(1)}$ or $x_{(2)} < x$, 
we can set 
$\det (x - \phi) = \exp\left({\rm tr}\,{\rm ln} (x-\phi)\right)$ 
unambiguously. We hen obtain   
\begin{equation} 
 \left\langle \det (x-\phi)^2 \right\rangle
 = 
 \exp 
 \left[ 
  2\,
  \left\langle 
   {\rm tr}\,{\rm ln} (x - \phi) 
  \right\rangle_{\rm c}  
   + 
  \frac{1}{2}\, 
  \left\langle 
   \left(2\,{\rm tr}\,{\rm ln} (x-\phi)\right)^2 
  \right\rangle_{\rm c} 
  + 
  \cdots 
 \right] \, ,
  \label{eq:detOutsideCutStart}
\end{equation}
where the subscript ``c'' 
denotes the connected Green function. 
Here, we have omitted tentatively the superscript ``0\text{-inst}''. 
 We note that 
\begin{eqnarray} 
 \left\langle {\rm tr}\,{\rm ln} (x - \phi)\right\rangle_{\rm c}  
 &=& 
 \left\langle 
  {\rm tr}\,{\rm ln} (x-\phi) 
 \right\rangle_{\rm disk} 
 + 
 {\cal O} 
 \left(\frac{1}{N}\right) \, , 
  \nonumber \\ 
 \left\langle 
  \left( {\rm tr}\,{\rm ln} (x-\phi) \right)^2 
 \right\rangle_{\rm c} 
 &=& 
 \left\langle 
  ({\rm tr}\,{\rm ln} (x - \phi))^2 
 \right\rangle_{\rm cylinder} 
 +{\cal O}\left(\frac{1}{N^2}\right) \, , 
\end{eqnarray}
where the subscripts ``disk'' and ``cylinder'' 
indicate the disk and cylinder amplitudes, respectively. 
 Thus, Eq.~(\ref{eq:detOutsideCutStart}) takes the form 
\begin{eqnarray} 
 \lefteqn{ 
  \left\langle \det (x - \phi)^2 \right\rangle e^{-NV(x)} 
 } \nonumber \\
 &=& 
 \exp 
 \left[ 
  N\, 
  \left( 
   2\, 
   \left\langle 
    \frac{1}{N}\,{\rm tr}\,{\rm ln} (x - \phi) 
   \right\rangle_{\rm disk} 
   - 
   V(x) 
  \right) 
  + 
  2\, 
  \left\langle (\tr\log (x-\phi))^2 \right\rangle_{\rm cylinder} 
  + 
  {\cal O}\left(\frac{1}{N}\right) 
 \right] \nonumber \\
 &=& 
 \exp 
 \left( 
  - N\,V_{\rm eff}^{(0)}(x) 
  - V_{\rm eff}^{(1)}(x) 
  + {\cal O}\left(\frac{1}{N}\right) 
 \right) \, .
\end{eqnarray} 
 The ${\cal O}(N)$-contribution $[-N\,V_{\rm eff}^{(0)}(x)]$ 
consists of $V_{\rm eff}^{(0)}(x)$ in Eq.~(\ref{eqn:Veff0}) 
(as can be checked by differentiation), 
while the ${\cal O}(N^0)$-contribution $V_{\rm eff}^{(1)}(x)$ is 
given by 
\begin{eqnarray}
 V_{\rm eff}^{(1)}(x) 
 &=& 
 - 2\, 
 \left\langle \left({\rm tr}\,{\rm ln} (x - \phi)\right)^2 
 \right\rangle_{\rm cylinder} 
  \nonumber \\ 
 &=& 
 - 2\, 
 {\rm ln} 
 \left( 
  1 
  + 
  \frac{x - \displaystyle{\frac{x_{(1)} + x_{(2)}}{2}}} 
       {\sqrt{(x - x_{(1)})(x - x_{(2)})}} 
 \right) 
 +2\, {\rm ln} 2 \, . 
\label{eqn:Veff1} 
\end{eqnarray} 
 In deriving Eq.~(\ref{eqn:Veff1}), 
we have invoked the formula for the cylinder amplitude 
\cite{Ambjorn:1990ji} 
\begin{eqnarray} 
 && 
 \left\langle 
  {\rm tr} 
  \left( \frac{1}{x - \phi} \right) 
  {\rm tr} 
  \left( \frac{1}{y-\phi} \right) 
 \right\rangle_{\rm cylinder} 
  \nonumber \\ 
 && \qquad 
 = 
 \frac{1}{2(x-y)^2} 
 \left( 
  \frac{\displaystyle{ 
         x y 
         - \frac{1}{2}(x_{(1)} + x_{(2)})(x + y) + x_{(1)} x_{(2)} 
        }} 
       {\sqrt{(x - x_{(1)}) (x - x_{(2)}) 
              (y - x_{(2)}) (y - x_{(1)})}} 
        - 1 
 \right) \, .            
\end{eqnarray} 
 Thus, the cylinder amplitude is given 
solely by the two endpoints $x_{(1)} $ and $ x_{(2)}$ 
where $x_{(1)} < x_{(2)}$, 
of the eigenvalue distribution, 
and it does not depend on any other details 
of the profile of the potential $V(x)$. 

 To evaluate the denominator of Eq.~(\ref{eqn:chempot}), 
it is also necessary to evaluate 
$V_{\rm eff}^{(1)}(x)$ for $x \in [x_{(1)},\,x_{(2)}]$. 
 However, Eq.~(\ref{eqn:Veff1}) is invalid in this region. 
 For instance, it diverges at $x = x_{(1)}$. 
 This implies 
that we need to elaborate on the $N$-dependence 
of the cylinder amplitude. 
It is necessary to perform the calculation 
up to the magnitude of $V_{\rm eff}^{(1)}$, 
that is, the next leading order in $\frac{1}{N}$.
If $V_{\rm eff}^{(1)}(x)$ in this region takes the form 
\begin{equation} 
 V_{\rm eff}^{(1)}(x) 
 = 
 - {\rm ln}\,N 
 + [\mbox{finite function of } x ] \, , 
  \label{eq:V1_expected}
\end{equation} 
the chemical potential becomes finite, 
because the $N$ dependence in Eq.~(\ref{eq:V1_expected}) 
cancels the overall factor $N$ 
in Eq.~(\ref{eqn:chempot}).  The goal of the rest of this section 
is to demonstrate 
that this is indeed the case 
by evaluating the right-hand side of Eq.~(\ref{eqn:chempot}) 
explicitly, using the orthogonal polynomial method. 
 As we see below, 
$\left<\det (x - \phi)^2\right>$ 
has a compact expression 
in terms of orthogonal polynomials. 
 The task at hand then reduces to evaluating 
the explicit forms of orthogonal polynomials 
in the regions $x \in [x_{(1)},\,x_{(2)}]$, 
$x < x_{(1)}$ and $x > x_{(2)}$ separately. 
 It turns out that 
the case $x \in [x_{(1)},\,x_{(2)}]$ 
requires a more involved analysis, 
for which we employ a sort of WKB approximation explicitly 
(where $\hbar = \frac{1}{N}$) 
in solving the recursion relation for the orthogonal polynomials. 
 By combining these ingredients, 
we finally obtain the chemical potential $\mu$ 
and show that $\mu$ is a ``universal'' quantity, 
independent of the details of the potential $V(x)$. 

\subsection{Orthogonal polynomials 
           and \mbox{\boldmath $\left<\det (x - \phi)^2\right>$} 
           } 
 It is known 
that a set of polynomials $P_n(x) = x^n + O(x^{n-1})$ 
($n \in {\bf Z}_+ \cup \{0\}$) 
that obeys the orthogonality condition 
\begin{equation} 
 \int 
  dx\, e^{-NV(x)}\, 
  P_n(x)\,P_m(x) 
 = h_n\,\delta_{nm} \, 
\label{eqn:orthogonality}  
\end{equation} 
can be expressed as the expectation value of $\det(x - \phi_{(n)})$ 
for a system 
of an  $n \times n$ Hermitian matrix $\phi_{(n)}$ 
with the action $N {\rm tr}\,V(\phi_{(n)})$ 
\cite{Eynard:1993fp}, 
\begin{equation} 
 P_n(x) 
 = 
 \frac{1}{Z_n}\, 
 \int d\phi_{(n)}\, 
  e^{-N\,{\rm tr}\,V(\phi_{(n)})}\, 
  \det(x - \phi_{(n)}) \, , \label{eq:Pndet}
\end{equation} 
where $Z_n$ is the partition function of that system. Note
the coefficient $N$ multiplying the potential $V(\phi_{(n)})$  in the
above expression. Equation (\ref{eq:Pndet}) leads us to seek an
expression of $\left<\det (x - \phi)^2\right>$ in terms of the
orthogonal polynomials, 
for example, something like $\left(P_N(x)\right)^2$. 
As a matter of fact,  $\left<\det (x - \phi)^2\right>$ 
does have   an explicit expression in terms of orthogonal polynomials 
and the coefficients of the recursion relation satisfied by them. 
We now demonstrate this point.
 
 Let us first define a new quantity $D_n(x)$ as
\begin{equation} 
 D_n(x) 
\equiv
 \det_{0 \le i,\,j \le n-1} (M_{ij}(x)) \, , 
\end{equation}
where 
\begin{equation}
 M_{ij}(x) 
 \equiv 
 \int dx^\prime\,e^{-NV(x^\prime)}\, 
 \sum_{k=0}^\infty 
  \frac{1}{\sqrt{h_i h_k}} P_i(x^\prime) 
   (x - Q)^2_{jk} 
   P_k(x^\prime) \, .
\end{equation} 
 Thus, 
$D_n(x)$ is the determinant of the $n \times n$ matrix 
consisting of the first $n\times n$ 
entries of the matrix $\{M_{ij}(x)\}_{i,\,j \ge 0}$. 
 In the above, 
$Q_{i,j}$ is defined 
through the recursion relation for the orthogonal polynomials 
$\{P_n\}_{n \in {\bf Z}_+ \cup \{0\}}$, 
\begin{equation} 
 x P_n(x) 
 = \sum_{m=0}^\infty Q_{n,m} P_m(x) \, .
\end{equation} 
 In particular, for the $n=N$ case we have 
\begin{equation}
 D_N(x) = \left<\det (x - \phi)^2\right> \, . 
\end{equation} 
 It is easy 
to see that $D_n(x)$ satisfies the recursion relation 
\begin{equation}
 D_n(x) 
 = 
 P_n^2(x) + Q_{n,n-1}\,D_{n-1}(x) \, . 
\end{equation} 
 It is also straightforward to show that  
$Q_{n,n+1} = 1$, while 
\begin{eqnarray} 
 && 
 Q_{n,n-1} = r_n \, , 
  \nonumber \\ 
 && 
 r_n \equiv \frac{h_n}{h_{n-1}} \ (n \ge 1)\, , 
 \quad 
 r_0 \equiv 0 \, .
\label{eqn:hratio}   
\end{eqnarray} 
 Repeated use of the above recursion relation leads to 
\begin{equation}
 D_N(x) 
 = 
 \left(P_N(x)\right)^2 
 + 
 r_N\,\left(P_{N-1}(x)\right)^2 
 + 
 \cdots 
 + 
 r_N \cdots r_1\,\left(P_0(x)\right)^2\, . 
  \label{eqn:DN} 
\end{equation} 
 This formula is the key equation 
in the evaluation of $\left<\det (x - \phi)^2\right>$, and it enables us  
to evaluate the chemical potential by using the asymptotic behavior of the orthogonal polynomials.  
 The remaining task is 
to evaluate the orthogonal polynomials $P_n(x)$ 
and the coefficients $r_n$ appearing in the recursion relation 
up to the next-to-leading order in the $\frac{1}{N}$ expansion. 
This is done in the following subsection.

\subsection{Asymptotic behavior of \mbox{\boldmath $P_n$}} 
\label{subsec:asmbehP}

 The forms of the orthogonal polynomials 
in the large-$N$ limit have been determined 
as follows \cite{Brezin:1993qg,Eynard:1997qu}: 
\begin{equation} 
 P_n(x) 
 = 
 \sqrt{\frac{2}{\pi}}\, 
 e^{\frac{N}{2}\, V(x)}\, 
 \frac{1}{\sqrt{(x - x_{(1)}) (x - x_{(2)})}}\, 
  \cos
  \left[ 
    N\,\chi_a(x) 
    - (N - n) \,\chi_b(x) 
    + \chi_c(x) 
  \right] \, .
\end{equation} 
The values $\chi_a, \chi_b$ and $\chi_c$ are evaluated 
in Refs.~\cite{Brezin:1993qg}, and \cite{Eynard:1997qu}.
Here, let us instead proceed to evaluate them directly 
from the recursion relation 
\begin{equation}
 x P_n(x) 
 = 
 P_{n+1}(x) 
 + s_n P_n(x) 
 + r_n P_{n-1} \, , 
 \label{Pzenkasiki} 
\end{equation} 
where $s_n = Q_{n,n}$ in terms of $Q_{n,m}$, in order to determine the
behavior of the orthogonal polynomials to next-to-leading order in 
${\cal O}\left(\frac{1}{N}\right)$.

%
 At leading order in $\frac{1}{N}$, 
$P_n(x)$ depends on $N$ as 
$ e^N$. Hence, it is convenient to consider the ratio 
\begin{equation} 
 k_n(x) 
 \equiv 
 \frac{P_n(x)}{P_{n-1}(x)} 
 \sim {\cal O}(N^0) \, , 
\end{equation} 
and to expand the recursion relation (\ref{Pzenkasiki}) 
in $\frac{1}{N}$. 
 With respect to $k_n(x)$, Eq.~(\ref{Pzenkasiki}) takes the form 
\begin{equation} 
 x 
 = 
 k_{n+1}(x) + s_n + \frac{r_n}{k_n(x)} \, . \label{Gzenkasiki}
\end{equation} 
 We expand $k_n(x)$ in $\frac{1}{N}$ as 
\begin{equation} 
 k_n(x) 
 = 
 k^{(0)}\left(x,\,\frac{n}{N}\right) 
 + 
 \frac{1}{N}\,k^{(1)}\left(x,\,\frac{n}{N}\right) 
 + {\cal O}\left(\frac{1}{N^2}\right)\, . 
  \label{eq:q_n_expansion} 
\end{equation} 

When we take $n$ and $N$ to the infinity and introduce the continuum variable $\xi=\frac{n}{N}$,
the leading and next-to-leading orders 
of the recursion relation (\ref{Gzenkasiki}) yield 
\begin{eqnarray} 
 && 
 \left( k^{(0)}(x,\,\xi) \right)^2 
 + (s(\xi) - x) k^{(0)}(x,\,\xi) 
 + r(\xi) 
 = 0 \, , 
  \nonumber \\ 
 && 
 \left( 
  2 k^{(0)}(x,\,\xi) + (s(\xi) - x) 
 \right) 
  k^{(1)}\left(x,\,\xi\right) 
 + 
 k^{(0)}(x,\,\xi) 
 \left( 
  \frac{1}{2}\,\partial_\xi s(\xi) 
  + 
  \partial_\xi k^{(0)}(x,\,\xi) 
 \right) 
 = 0  \, , 
  \nonumber \\ 
\end{eqnarray} 
respectively. Here, the continuum limits of $r$ and $s$ are taken 
as expressed in Eq. (\ref{eq:rs_r()s()}) (see Appendix C).
 By solving these equations, we find 
\begin{equation} 
 k^{(0)}(x,\,\xi) 
 = 
 \frac{- (s(\xi) - x) 
     \pm \sqrt{(s(\xi) - x)^2 - 4 r(\xi)}}{2} 
 \, ,  
  \label{eq:q0} 
  \end{equation}
 where the branch of the square root should be chosen so that $k_n(x) \sim x$ as $|x| \rightarrow \infty$ 
by taking account of the original definition of $k_n(x)$. We also find
\begin{equation}
 k^{(1)}(x,\,\xi) 
 =
 - 
 \frac{k^{(0)}(x,\,\xi) 
       \left( 
        \frac{1}{2}\,\partial_\xi s(x,\,\xi) + \partial_\xi k^{(0)}(x,\,\xi) 
       \right)} 
      {2 k^{(0)}(x,\,\xi) + s(\xi) - x} .
  \label{eq:q1}
\end{equation} 
The expression for $k^{(1)}(x,\,\xi)$ 
can be further simplified as 
\begin{equation}
 k^{(1)}(x,\,\xi) 
 =
- \frac{1}{2}\,k^{(0)}(x,\,\xi) 
 \partial_\xi\,{\rm ln}\,q(x,\,\xi) 
  \, , 
  \label{eq:q1byu}
\end{equation} 
where 
\begin{eqnarray} 
 q(x,\,\xi) 
 &=& 
 2 k^{(0)}(x,\,\xi) + s(\xi) - x 
  \nonumber \\ 
 &=& 
\pm \sqrt{\left(s(\xi) - x\right)^2 - 4 r(\xi)} \, . 
\label{eq:u} 
\end{eqnarray} 

Now, it is straightforward 
to write down the expression for $P_n(x)$ as a product of the $k_i(x)$: 
\begin{eqnarray} 
 P_n(x) 
 &=& 
 \prod_{j=1}^n k_j(x) 
=
 \exp 
 \left[ \sum_{j=1}^n {\rm ln}\,k_j(x) \right]  
  \nonumber \\ 
 &=& 
 \exp 
 \left[ 
  N \int_0^{\frac{n}{N}} d\xi\,{\rm ln}\,k^{(0)}(x,\,\xi) 
  +  \int_0^{\frac{n}{N}} d\xi\, 
  \frac{k^{(1)}(x,\,\xi)}{k^{(0)}(x,\,\xi)} 
\right. 
  \nonumber \\ 
 && \qquad 
 \left. 
  + 
  \frac{1}{2} 
  \left( 
   {\rm ln}\,k^{(0)}\left(x,\,\frac{n}{N}\right) 
   - 
   {\rm ln}\,k^{(0)}\left(x,\,0\right)
  \right) 
   + {\cal O}\left(\frac{1}{N}\right)
  \right] \, , 
  \label{eq:expr_Pn_tent}
\end{eqnarray} 
where we have utilized 
the following form 
of the Euler-Maclaurin summation formula 
to convert the summation into an integral 
\begin{equation} 
 \sum_{j = n_1}^{n_2} f_n 
 = 
 N \int_{\frac{n_1}{N}}^{\frac{n_2}{N}} d\xi\,f(\xi) 
 + 
 \frac{1}{2} 
 \left( 
  f\left(\frac{n_2}{N}\right) - f\left(\frac{n_1}{N}\right) 
 \right) 
 + 
 {\cal O}\left(\frac{1}{N}\right) \, . 
\end{equation} 
 Using the expression (\ref{eq:q1byu}) for $k^{(1)}(x)$ and 
the facts that $r(0) = 0 = s(0)$ 
and $q(x,0)=x=k^{(0)}(x,0)$, 
we obtain an expression for the asymptotic behavior 
of the orthogonal polynomials as 
\begin{eqnarray} 
 P_n(x) 
 &=& 
 \exp 
 \left[ 
  N \int_0^{\frac{n}{N}} d\xi\,{\rm ln}\,k^{(0)}(x,\,\xi) 
  + 
  \frac{1}{2}\,  
   {\rm ln}\,k^{(0)}\left(x,\,\frac{n}{N}\right) 
  - 
  \frac{1}{2}\,  
   {\rm ln}\,q\left(x,\,\frac{n}{N}\right) 
 \right. 
  \nonumber \\ 
 && \qquad 
 \left. 
  + {\cal O}\left(\frac{1}{N}\right)
 \right] 
  \nonumber \\ 
 &=& 
 \left( 
  \frac{k^{(0)}\left(x,\,\frac{n}{N}\right)} 
       {q\left(x,\,\frac{n}{N}\right)} 
 \right)^{\frac{1}{2}} 
 \, 
 \exp 
 \left[ 
  N \int_0^{\frac{n}{N}} d\xi\,{\rm ln}\,k^{(0)}(x,\,\xi) 
 \right] 
 \left( 
  1 + {\cal O}\left(\frac{1}{N}\right)
 \right) \, ,
  \label{eq:asymptotic_Pn}
\end{eqnarray} 
where $k^{(0)}(x)$ and $q(x,\,\xi)$ 
are given explicitly by Eqs.~(\ref{eq:q0}) and (\ref{eq:u}), 
respectively. 

%
Thus, we have obtained an expression for $P_n(x)$, (\ref{eq:asymptotic_Pn}), 
for a given (real) value of $x$ by solving this equation in terms of $n$. 
 However, the above expression becomes complex 
when the argument of the square root in $q(x,\,\xi)$ is negative, i.e.,
when  
\begin{equation} 
 \left( 
  x - s\left(\frac{n}{N}\right) 
 \right)^2 
 - 4\,r\left(\frac{n}{N}\right) 
 < 0 \, .
\end{equation} 
 This seems to be a contradiction, 
because $P_n(x)$ should be a real polynomial. 
 To resolve this difficulty, 
let us consider an actual plot of $P_n(x)$. 
There is a finite region 
in which it exhibits oscillating behavior with $n$ nodes. 
This oscillatory region 
precisely corresponds to the region  
where the exponent of $P_n(x)$ 
becomes imaginary in the expression (\ref{eq:asymptotic_Pn}), 
because $P_n(x)$ contains $q(x,\,\xi)$ in its exponent. 
 This suggests that 
in this oscillatory region, $P_n(x)$ in Eq.~(\ref{eq:asymptotic_Pn}) 
should behave 
like a trigonometric function. 

 Now, it is helpful 
to recall that we have derived the expression (\ref{eq:asymptotic_Pn}) 
by starting from a set of linear difference equations. 
 For this derivation, 
we can draw an analogy 
with the WKB method in usual quantum mechanics. 
 The important condition here is 
the same as the continuity condition for the solution 
at the turning point, 
where the behavior of the wavefunction changes
from decaying behavior in the classically forbidden region 
to the oscillating behavior in the classically allowed region. 
 Thus, we should multiply the naive expression in Eq.~(\ref{eq:asymptotic_Pn}) 
by an appropriate phase factor and take its real part. 
 In other words, 
the phase factor emerging in the analytic continuation 
of Eq.~(\ref{eq:asymptotic_Pn}) 
should be combined to make $P_n(x)$ 
a real trigonometric function with an appropriate phase. 

 In Appendix D, we show that exactly the same argument as for the WKB approximation applies here
and, in particular, we should take an extra factor of $2$ into account  
when Eq.~(\ref{eq:asymptotic_Pn}) is continued 
into the oscillating region analytically. Noting also that 
$\left|k^{(0)}(x,\xi)\right| = \sqrt{r(\xi)}$, 
we obtain the following expression for $P_n(x)$ 
in the oscillating region: 
\begin{eqnarray} 
 P_n(x) 
 &=& 
 2\left( 
 \frac{\sqrt{r\left( \frac{n}{N}\right)}} 
      {\left|q\left(x,\,\frac{n}{N}\right)\right|} 
 \right)^{\frac{1}{2}}  
 \, 
 \exp 
 \left[ 
  N\,{\rm Re}\, 
  \int_0^{\frac{n}{N}} d\xi\,{\rm ln}\,k^{(0)}(x,\,\xi) 
 \right] 
 \sin\left(\theta(x)\right) 
  \nonumber \\ 
 && \quad 
 \times 
 \left( 
  1 + {\cal O}\left(\frac{1}{N}\right)
 \right) \, . 
  \label{eq:Pn_allowed_region}
\end{eqnarray} 
Here $\theta(x)$ represents a phase factor. 
As there are $n \sim {\cal O}(N)$ nodes 
located in a finite interval, 
$\theta(x)$ is expected to vary rapidly, 
while Eq.~(\ref{eq:Pn_allowed_region}) 
explicitly shows that 
the amplitude of the oscillation changes very slowly. 
For our purposes, it suffices to know the averaged value  taken over a small interval around $x$. 
Taking the average of the square of $P_n(x)$ for an interval where $\theta(x)$ should oscillate at least once, we obtain 
\begin{eqnarray}
 \left(P_{n}(x)\right)^2 _{\rm averaged}
 &=&  
 2\, 
  \frac{\sqrt{r\left( \frac{n}{N}\right)}} 
       {\left|q\left(x,\,\frac{n}{N}\right)\right|} 
   \, 
 \exp 
 \left[ 
  2 N\,{\rm Re}\,\int_0^{\frac{n}{N}} d\xi\,{\rm ln}\,k^{(0)}(x,\,\xi) 
 \right] 
  \nonumber \\ 
 && \quad 
 \times 
 \left( 
  1 + {\cal O}\left(\frac{1}{N}\right) 
 \right)
 \, ,
  \label{Psquare}
\end{eqnarray} 
which we use hereafter as the  asymptotic form of the orthogonal polynomials when $x$ lies in the oscillating region.

\subsection{Evaluation of \mbox{\boldmath$\left<\det (x - \phi)^2\right>$} }
\label{subsec:evalofchar.poly.}
Now we are ready to evaluate $\left<\det (x - \phi)^2\right>$ using the
asymptotic behavior of the orthogonal polynomials obtained in the
previous section. There is an essential difference between the case in
which $x$ is positioned inside the cut, that is, $x_{(1)}<x<x_{(2)}$, and
the case in which $x$ is outside the cut.

Let us first consider the case that $x$ is located outside  the cut.
In (\ref{eqn:DN}), the ratio of neighboring terms, which we write as $ c_j(x) $, is
\begin{eqnarray} 
 c_j(x) 
 &\equiv& 
 \frac{\left(P_{N-j-1}(x)\right)^2 r_{N-j}} 
      {\left(P_{N-j}(x)\right)^2} 
=
 \frac{r_{N-j}}{\left(k_{N-j}(x)\right)^2} 
  \nonumber \\ 
 &=& 
 \frac{r\left(\frac{N-j}{N}\right)} 
      {\left(k^{(0)}\left(x,\,\frac{N-j}{N}\right)\right)^2} 
 + {\cal O}\left(\frac{1}{N}\right) 
  \nonumber \\ 
 &=& 
 \left. 
  \frac{-(s(\xi) - x) - \sqrt{(s(\xi) - x)^2 - 4 r(\xi)}} 
       {-(s(\xi) - x) + \sqrt{(s(\xi) - x)^2 - 4 r(\xi)}} 
 \right|_{\xi = \frac{N-j}{N}} 
 + {\cal O}\left(\frac{1}{N}\right) \, .
\end{eqnarray} 
The quantity $c_j(x)$ is of ${\cal O}(N^0)$ and less than 1. 
Therefore, the largest contribution to $\left<\det (x - \phi)^2\right>$
comes from the first term $P_N(x)^2$. Note also that the 
dependence of $c_j(x)$ on $j$ is modest 
[i.e. ${\cal O}(N^0)$] compared to the number [${\cal O}(N)$] of terms. 
 Hence, when the sum is replaced by a sum of 
terms with a constant ratio, say $c_0(x)$, 
the correction will be of order $\frac{1}{N}$: 
\begin{eqnarray} 
 \left< \det (x - \phi)^2 \right>
 &=& 
 \left(P_N(x)\right)^2 
 \sum_{j=0}^N \left(c_0(x)\right)^j 
 \left( 1 + {\cal O}\left(\frac{1}{N}\right) \right) 
  \nonumber \\ 
 &=& 
 \left(P_N(x)\right)^2\, 
 \frac{1}{1 - c_0(x)} 
 \left( 1 + {\cal O}\left(\frac{1}{N}\right) \right) \, .  \label{eqn:det1}
\end{eqnarray} 
The factor $\frac{1}{1 - c_0(x)}$ turns out to be 
\begin{eqnarray} 
 \frac{1}{1 - c_0(x)} 
 &=& 
 \frac{-(s(1) - x) + \sqrt{(s(1) - x)^2 - 4 r(1)})} 
      {2 \sqrt{(s(1) - x)^2 - 4 r(1)}} 
  \nonumber \\ 
 &=& 
 \frac{k^{(0)}(x,\,1)}{q(x,\,1)} \, .
\end{eqnarray} 
Hence, we obtain  
 \begin{equation} 
 \left< \det (x - \phi)^2 \right>
= 
 \left( 
  \frac{k^{(0)}\left(x,\,1\right)} 
       {q\left(x,\,1\right)}  \right)^2 
\exp 
 \left[ 
  2 N \int_0^1 d\xi\, 
  {\rm ln}\, k^{(0)}\left(x,\,\xi\right) 
 \right] 
 \left( 
  1 + {\cal O}\left(\frac{1}{N}\right)
 \right) .
\end{equation}

Next, we evaluate $\left<\det (x - \phi)^2\right>$ for the case that $x$
lies inside the cut, that is, in $[x_{(1)},x_{(2)}]$.
Note first, as briefly described in the previous subsection, the
orthogonal polynomials $P_n(x)$ have the following oscillatory behavior:
(1) $P_n(x)$ oscillates within a finite region in which it has $n$ nodes. 
The amplitude of each oscillation itself varies
slowly, so that we can draw a curve enveloping these
oscillations. Outside this oscillatory region, $P_n(x)$ behaves 
as  $x^n$. (2) This oscillatory region grows as $n$ becomes large. 
For sufficiently large $n$, this region becomes the region in which $k^{(0)}$
becomes imaginary,
\begin{equation}
\left( x-s\left(\frac{n}{N}\right)\right)^2 
-4r\left(\frac{n}{N}\right) <0.
\end{equation}
Thus, the main contribution to $D_N$ comes 
from the terms contains $P_m(x)$ with $m>n_0$, where the integer $n_0$ 
for a given $x$ is defined by
\begin{equation}
\left( x-s\left(\frac{n_0(x)}{N}\right)\right)^2 
-4r\left(\frac{n_0(x)}{N}\right) =0.
\end{equation}
We can repeat the argument that leads to (\ref{eqn:det1}) for this case. 
We can thereby show that the contribution from $P_m(x)$ for $m<n_0(x)$
is suppressed exponentially, because the ratio of the neighboring terms
is less than 1.
Hence, $\left<\det (x - \phi)^2\right>$ 
can be approximated as  
\begin{equation} 
 \left<\det (x - \phi)^2\right>
 = 
 \sum_{i= n_0(x)}^N 
 \left(P_i(x)\right)^2 
 \prod_{j=0}^{N-1-i} r_{N-j} \, . 
  \label{eq:det_truncated} 
\end{equation} 
From the expression for $ \left(P_i(x)\right)^2$ in (\ref{Psquare}) and
an additional calculation presented in Appendix E, we obtain
\begin{eqnarray} 
\left<\det (x - \phi)^2
 \right>_{\rm averaged}
 &=& 
2N\, 
 \pi\,\sqrt{r(1)}\,\rho(x)\, 
 \exp 
 \left[ 
  2N 
  \int_{0}^{\xi_0(x)} d\xi\, {\rm ln}\,k^{(0)}(\xi)   + N 
  \int_{\xi_0(x)}^{1} d\xi\, {\rm ln}\,r(\xi) 
 \right] 
  \nonumber \\ 
 && \quad 
 \times 
 \left( 
  1 + {\cal O}\left(\frac{1}{N}\right)
 \right) \, , \label{eqn:detxphiaveraged}
\end{eqnarray} 
where $\xi_0(x)=\frac{n_0(x)}{N}$ and $\rho(x)$ is the eigenvalue density (\ref{eq:rhodef}).
%
%

We summarize the above result in the form of the effective potential $V_{\rm eff}$ from the relation 
$e^{-N(V_{\rm eff}(x)-V(x))}= \left<\det (x - \phi)^2\right>$.
\begin{eqnarray}
\mbox{\rm   Outside the cut:} \qquad V_{\rm eff}(x)  &=  & V(x) -2 \int^1_0 d\xi {\rm ln} k^{(0)}(\xi) \nonumber \\
&& -\frac{1}{N}\Big\{ -\log [ (x-s(1))^2-4r(1) ] \label{veffout} \\
&&+2\log \left( x-s(1)+\sqrt{(x-s(1))^2-4r(1)} \right)-2\log 2 \Big\} . \nonumber \\
\mbox{\rm   Inside the cut:} \qquad V_{\rm eff}(x) &=&  V(x) -2 \int_{0}^{\xi_0(x)} d\xi\, {\rm ln}\,k^{(0)}(\xi) -  \int_{\xi_0(x)}^{1} d\xi\, {\rm ln}\,r(\xi) \nonumber \\ && \qquad 
-\frac{1}{N}\log\left[ 2N\pi\rho(x)\sqrt{r(1)} \right]. \label{veffin}
\end{eqnarray}
Also, if we use $V_{\rm eff}^{(0)}$ in Eq.~(\ref{eqn:Veff0}), 
the above expressions can be cast into much simpler forms.
\begin{eqnarray}
\mbox{\rm   Outside the cut:} \qquad V_{\rm eff}(x)  &=  & V_{\rm eff}^{(0)}(x) -\frac{1}{N}\left\{ -2\log \left[\frac{k^{(0)}(x,1)}{q(x,1)}\right]\right\}. \label{veffout2} \\
\mbox{\rm   Inside the cut:} \qquad V_{\rm eff}(x) &=& V_{\rm eff}^{(0)}(x)-\frac{1}{N}\log\left[ 2N\pi\rho(x)\sqrt{r(1)} \right]. \label{veffin2}
\end{eqnarray}

\subsection{Universality of the chemical potential of the instanton}
Now we are ready to take the ratio of the partition functions 
existing inside and outside the cut. They are given by the integral of 
$V_{\rm eff}$ above obtained for the respective regions. We evaluate
this ratio, the chemical potential of the instanton, near the critical
point.

We can evaluate the contribution from outside the cut using the saddle
point method. To do this, let us first evaluate the leading contribution 
to $V_{\rm eff}$, that is the first two terms on the right-hand side of
(\ref{veffout}),
$$
 V(x) -2 \int^1_0 d\xi {\rm ln} k^{(0)}(\xi).
$$
Near the critical point, as explained in Appendix C, $r$ and $s$ behave as 
\begin{eqnarray}
r&=&r_c -\frac12\alpha r_c \sqrt{1-\xi +\Delta \kern -1.5pt g } \label{rcrit}, \\
s&=&s_c - \frac12\alpha \sqrt{r_c}\sqrt{1-\xi +\Delta \kern -1.5pt g },
\end{eqnarray}
where $\Delta \kern -1.5pt g$ represents the deviation from the critical point and $\alpha$ is a certain constant.
If we rescale $x$ and $\xi$ to emphasize the region near the critical point as 
\begin{eqnarray}
x&=&s_c+2\sqrt{r_c}+\sqrt{\Delta \kern -1.5pt g}\zeta ,\\
\xi &=&1-\Delta \kern -1.5pt g \eta,
\end{eqnarray}
$V_{\rm eff}$ can be expanded in terms of $\Delta \kern -1.5pt g$ as
\begin{equation}
V_{\rm eff} \sim V(x) -\int_0^{\Delta \kern -1.5pt g^{-1}} d\eta \Delta \kern -1.5pt g  \left( 
2\ln(\sqrt{r_c})+\Delta \kern -1.5pt g^{\frac14}\frac{2\sqrt{\zeta\sqrt{r_c}+r_c \alpha \sqrt{\eta+1}}}{\sqrt{r_c}} + {\cal O}(\Delta \kern -1.5pt g ^\frac12) \right) .
\label{veffexpand}
\end{equation}

We are interested in the universal part of the derivative of $V_{\rm
eff}$ rather than $V_{\rm eff}$ itself. 
In other words, we are interested in the term proportional to $\Delta \kern -1.5pt g ^\frac54$, which can be obtained in the following familiar form after integrating over $\eta$:
\begin{equation}
V'_{\rm eff}(\zeta) \sim -\frac83 \Delta \kern -1.5pt g ^\frac54 \frac{\left(\zeta-\frac12\sqrt{r_c}\alpha\right) \sqrt{\zeta+\sqrt{r_c}\alpha}}{r_c^\frac54 \alpha^2}. \label{vprime}
\end{equation}
 Thus, at leading order, the integral $\int dx e^{-V_{\rm eff}}$ can be
 evaluated at the saddle point, $\zeta = \frac12\sqrt{r_c}\alpha$. At
 the next-to-leading order, that is, for the contribution from the cylinder part, we can use the same saddle point as that at leading order. The standard saddle point calculation yields
\begin{equation}
\int_{x<x_{(1)},x_{(2)}<x} dx e^{-NV_{\rm eff}(x)}=\frac{i}{12}\frac{3^\frac14 2^\frac14 \sqrt{\pi}\sqrt{r_c}}{\alpha^{\frac14}\Delta \kern -1.5pt g^\frac58 \sqrt{N}} e^{-NV_{\rm eff}^{(0)}(x_{0})}. \label{znum}
\end{equation}

The contribution from inside the cut is simpler to evaluate. 
Noting that  for this case, $V_{\rm eff}^{(0)}(x)$ takes a constant value, say  $V_{\rm eff}^{(0)}(x_{(2)})$, and we can replace $r_N$ with $r_c$,
it is straightforward to obtain
\begin{equation}
\int_{x_{(1)}<x<x_{(2)}} dx e^{-NV_{\rm eff}(x)}=\int_{x_{(1)}<x<x_{(2)}} dx 2N\pi\rho(x)\sqrt{r_N}e^{-NV_{\rm eff}^{(0)}(x)}=2N\pi\sqrt{r_c} e^{-NV_{\rm eff}^{(0)}(x_{(2)})}. \label{zden}
\end{equation}
In passing, we note that we can perform another  calculation that leads rather directly to (\ref{zden}). This calculation, which we present in Appendix E, does not involve the explicit form of $V_{\rm eff}(x)$. 

Taking the ratio of (\ref{znum}) and (\ref{zden}), we finally obtain the chemical potential of the instanton (\ref{eqn:chempot}) as
\begin{equation}
\mu= \frac{i}{24}\frac{3^\frac14 2^\frac14}{\alpha^{\frac14}\Delta \kern -1.5pt g^\frac58\sqrt{\pi} \sqrt{N}}e^{-N\left(V_{\rm eff}^{(0)}(x_{(0)})-V_{\rm eff}^{(0)}(x_{(2)})\right)}.
\label{chempot1}
\end{equation}

To make a connection with the standard analysis, we need to rewrite the chemical potential (\ref{chempot1}) in terms of a quantity that appears in the string equation.
The solution of the string equation is the second derivative of the free energy,  and it is conventionally normalized so that $\ddot{F}\sim -\sqrt{t}$. With this normalization, the free energy is given by 
\begin{equation}
-\frac{4}{15}t^\frac52. \label{Fstring}
\end{equation}
On the other hand, the free energy of the matrix model is obtained as
\begin{equation}
F_{\rm matrix}=N^2 \int^1_0(1-\xi)\log(r(\xi))d\xi.
\end{equation}
Near the critical point, $r$ behaves as (\ref{rcrit}). Expanding $F_{\rm matrix}$ in terms of $\Delta \kern -1.5pt g$ and taking the first non-integer exponent yields the universal part of the free energy as
\begin{equation}
-\frac{2N^2\alpha\Delta \kern -1.5pt g^\frac52}{15}. \label{Fmatrix}
\end{equation}
By comparing (\ref{Fstring}) and (\ref{Fmatrix}), we can connect $\Delta \kern -1.5pt g$ with $t$ in the string equation as 
\begin{equation}
\Delta \kern -1.5pt g = \frac{2^\frac25 t}{\alpha^\frac25 N^\frac45}.
\label{relation_g_t}
\end{equation}
Using this relation and invoking Eq. (\ref{eqn:Sinst}), we reach the universal expression for the chemical potential
\begin{equation}
\mu=\frac{Z_N^{({1\text{-inst}})}}{Z_N^{({0\text{-inst}})}}
=\frac{i}{8\cdot3^\frac34\sqrt{\pi}t^\frac58}
 e^{-\frac{8\sqrt3}{5}t^\frac54}.\label{universalmu}
\end{equation}

If $V(x)$ is an even function, the above argument needs two
modifications, due to the accidental symmetry of the potential with
respect to the exchange of  $x$ and $-x$.
First,  $s$ vanishes identically. 
However, \eqref{veffexpand}, \eqref{vprime}, \eqref{znum} and
\eqref{chempot1} hold if we replace $\alpha$ with $\frac\alpha 2$ in these equations.
Then \eqref{chempot1} becomes
\begin{equation}
\frac{i}{24}\frac{3^\frac14 2^\frac12}{\alpha^{\frac14}\Delta \kern -1.5pt g^\frac58\sqrt{\pi} \sqrt{N}}e^{-N\left(V_{\rm eff}^{(0)}(x_{(0)})-V_{\rm eff}^{(0)}(x_{(2)})\right)}.
\label{chempot4even}
\end{equation}
Secondly,  there emerge two critical points corresponding to two maximum
points of the potential that have exactly the same height. Each
critical point contributes the same amount to the free energy, and therefore 
the value of the free energy is twice that in generic potential
cases. 
Taking this effect into account, \eqref{Fstring} should be
\begin{equation}
-\frac{8}{15}t^{\frac52}.
\end{equation}
On the other hand, because $s$ does not contribute to the free energy, 
\eqref{Fmatrix} does not change.
Thus the relation between $\Delta g$ and $t$ becomes
\begin{equation}
\Delta g = \frac{2^{\frac45}t}{\alpha^\frac25N^\frac45}.
\end{equation}
Substituting this relation into \eqref{chempot4even},
we reach the same result. Hence, it is concluded that \eqref{universalmu} is universal.
As a concrete example, we present a computation 
of $\mu$ in the cases of both the $\phi^3$ and $\phi^4$ potentials in Appendix F. 


\section{D-instanton effect in loop equations} 
\label{sec:loop_eq} 
\setcounter{equation}{0}

 The Schwinger-Dyson equations
for the correlation functions of the ``loop variables'' 
${\rm tr}\,\phi ^{n}$ 
are known as the loop equations, and it is believed that they
give a complete description of the system.
Their continuum limit is easily taken \cite{loop_eq}, and 
the result can be interpreted as a sort of 
closed string field theory 
\cite{Ishibashi:1993pc,Ishibashi:1996er,Jevicki:1993rr}. 
 The loop variables are also useful when we compare the 
matrix model with the Liouville field theory. 

 In this section, we consider the question of what properties of the 
instantons observed in the previous sections 
can be captured by the continuum loop equations. 
 After some preparation in \S\ref{subsec:loop_eq}, 
we examine the classical limit
($g_{\rm st} \rightarrow 0$) of the loop equations
in \S\ref{subsec:classical},
and we see
that in the single-instanton vacuum,
the classical solution for the loop amplitude
has an ambiguity in the form of an arbitrary constant factor. 
 In \S\ref{subsec:beyond_classical}, 
we show that this constant factor can be determined if 
we consider the loop equations to all orders. 
 In \S\ref{subsec:loop_amp},  
we first show that the single-instanton loop amplitude obtained 
in this manner indeed is identical to 
that obtained from the matrix model calculation. 
 We next compute 
the D-instanton effect on the loop amplitude 
in the Liouville field theory, 
and we find that it too reproduces the result of the 
matrix model.  
 This coincidence confirms that 
the instanton in the matrix model 
is identical to the D-instanton in the Liouville field theory.

 To this point, we have seen that the loop equations correctly 
describe the loop amplitude in each vacuum with a fixed
number of instantons. 
 However, the continuum loop equations cannot 
determine the chemical potential of the instanton.  
 In \S\ref{subsec:cp_loop_eq}, 
we see that the continuum loop equations give a 
divergent expression and require regularization
when we attempt to evaluate the chemical potential.

\subsection{Loop equations} 
\label{subsec:loop_eq} 
 In this subsection, 
we make a few remarks on the loop equations 
that will be important in the succeeding analysis. 
 If we start with the matrix model with the action 
\begin{eqnarray} 
 && 
 S = N\,{\rm tr}\,V(\phi) \, , 
  \nonumber \\ 
 && 
 V(x) = \sum_{m=1}^p \frac{c_m}{m}\,x^m \, , 
\end{eqnarray} 
the loop equations can be derived from the Schwinger-Dyson equations 
\begin{equation} 
 0 = 
 \int 
 d\phi\, 
 \frac{\del}{\del \phi^a} 
 \left[ 
  {\rm tr}\left(t^a \phi^{n_0}\right)\, 
  W_{n_1} \cdots W_{n_r} 
  e^{-N\,{\rm tr}\,V(\phi)} 
 \right] \, . 
  \label{eq:SD_start}
\end{equation} 
 Here, $W_n$ is the loop variable 
\begin{equation} 
 W_n = \frac{1}{N}\,{\rm tr}\,\phi^n \, , 
\end{equation} 
and $\phi = \phi^a t^a$ with the $U(N)$ generator $t^a$ 
normalized so that ${\rm tr}(t^a t^b) = \delta^{ab}$. 
 Using 
\begin{eqnarray} 
 && 
 {\rm tr}(t^a A)\,{\rm tr}(t^a B) = {\rm tr}(AB) 
  \nonumber \\ 
 && 
 {\rm tr}(t^a A\, t^a B) = {\rm tr} A\, {\rm tr} B \, , 
\end{eqnarray} 
we can show that Eq.~(\ref{eq:SD_start}) takes the form 
\begin{eqnarray} 
 && 
 \sum_{j=0}^{n_0 - 1} 
 \left< 
  W_j W_{n_0 - 1 - j} W_{n_1} \cdots W_{n_r} 
 \right> 
 - 
 \sum_{m = 1}^{p} c_m 
  \left< 
   W_{n_0 + m - 1} W_{n_1} \cdots W_{n_r} 
  \right> 
  \nonumber \\ 
 && \quad 
 + 
 \frac{1}{N^2} 
 \sum_{s=1}^r n_s 
 \left< 
  W_{n_1} \cdots W_{n_{s-1}} W_{n_s - 1 + n_0} 
  W_{n_{s+1}} \cdots W_{n_r} 
 \right> 
 = 0 \, . 
  \label{eq:SD_eq_start}
\end{eqnarray} 
 We would like to find 
the form of Eq.~(\ref{eq:SD_eq_start}) 
in the double scaling limit, expressed by
\begin{eqnarray} 
 g 
 &=& g_\star\,e^{-a^2\,\beta\,t} \, , 
  \nonumber \\ 
 x &=& x_\star\,e^{a\,\gamma\,\zeta} \, , 
  \nonumber \\ 
 \frac{1}{N} 
 &=& a^{\frac{5}{2}}\,\kappa\,g_s \, , 
  \label{eq:double_scaling_limit} 
\end{eqnarray} 
for $\phi^3$-theory in which
$V(x) = \frac{1}{2}\,x^2 - \frac{g}{3}\,x^3$, 
and by
\begin{eqnarray} 
 g 
 &=& g_\star\,e^{-a^2\,\beta\,t} \, , 
  \nonumber \\ 
 x &=& x_\star\,e^{a\,\gamma\,\zeta} \, , 
  \nonumber \\ 
 \frac{1}{N} 
 &=& a^{\frac{5}{2}}\,\kappa\,g_s \, , 
  \label{eq:double_scaling_limit_phi4} 
\end{eqnarray} 
for $\phi^4$-theory in which
$V(x) = \frac{1}{2}\,x^2 - \frac{g}{4}\,x^4$. 
 In Eqs.~(\ref{eq:double_scaling_limit}) 
and (\ref{eq:double_scaling_limit_phi4}), 
$x_\star$ and $g_\star$ 
represent the critical values of $x$ and $g$, 
whose values, as well as those of $\beta$, $\gamma$ and $\kappa$,
are given explicitly in Appendices A and B 
for $\phi^3$-theory and $\phi^4$-theory, respectively. 
If we take the continuum loop length $l$ 
as $l = \gamma n a$, 
$W_n$ approaches the continuum loop operator 
$w(l)$ 
in the double scaling limit as 
\begin{equation} 
 W_n 
 \rightarrow 
 a^{\frac{5}{2}} x_\star^{n+1} \gamma Z_W^{\frac{1}{2}}\,w(l) \, , 
\end{equation} 
where the constant $Z_W$ is given by
\begin{equation} 
 Z_W = 6 \gamma^3 (g_\star x_\star)^3 
 \quad \mbox{for $\phi^3$-theory} \,  
\end{equation}   
 and
\begin{equation}  
 Z_W = \left(\frac{8}{3}\right)^2 \gamma^3 
 \quad \mbox{for $\phi^4$-theory} \, .         
  \label{eq:Z_Ws}
\end{equation}
 Using this, 
we can show that the loop equations 
(\ref{eq:SD_eq_start}) have the following 
continuum form in the double scaling limit
\cite{Ishibashi:1993pc}:
\begin{eqnarray} 
 && 
 l 
 \int_0^l dl^\prime 
 \left< 
  w(l^\prime)\,w(l - l^\prime)\,w(l_1) \cdots w(l_r) 
 \right> 
 + 
 \rho_s(l) 
 \left< 
  w(l_1) \cdots w(l_r) 
 \right> 
  \nonumber \\ 
 && \quad 
 + 
 g_{\rm st}^2\, l 
 \sum_{s=1}^r l_s 
 \left< 
  w(l_1) \cdots w(l_{s-1}) w(l_s + l) w(l_{s+1}) \cdots w(l_r) 
 \right> 
 = 0 \, , 
  \nonumber \\ 
 && 
 \rho_s(l) 
 \equiv 3\,\delta^{\prime\prime}(l) - \frac{3}{4}\,t \delta(l) \, . 
  \label{eq:loop_equations_w} 
\end{eqnarray} 

 We note that 
$g_{\rm st}$ appearing in Eq.~(\ref{eq:loop_equations_w}) 
differs from $g_s$ in the matrix model 
by the factor 
\begin{equation} 
 g_{\rm st} 
 = 
 \frac{1}{\sqrt{Z_W}}\, \frac{\kappa}{\gamma x_\star}\, g_s \, . 
\end{equation} 
 Using $Z_W$ in Eq.~(\ref{eq:Z_Ws}) 
and the values for $\beta$, $\gamma$ and $\kappa$ 
given in Appendices A and B, 
we find that 
the conversion factor 
$\frac{1}{\sqrt{Z_W}}\, \frac{\kappa}{\gamma x_\star}$ 
takes the same value 
in both $\phi^3$-theory and $\phi^4$-theory, 
\begin{equation} 
 g_{\rm st} 
 = 
 \frac{3}{4 \sqrt{2}}\,g_s \, . 
  \label{eq:g_S_translate} 
\end{equation} 
 We also remark that we employ such a convention that 
the same renormalized cosmological constant $t$ 
and renormalized boundary cosmological constant $\zeta$ 
can be used in the loop equations and the matrix model. 
 Therefore, we can compare the quantities in the loop equations 
with those in the matrix model 
by taking account of the conversion factor 
(\ref{eq:g_S_translate}) for the string coupling constants. 

 To treat the coupled equations (\ref{eq:loop_equations_w}), 
we introduce canonical pairs $\psi(l)$ and $\psi^\dagger(l)$ 
of closed string fields that satisfy the relations
\begin{eqnarray}  
 & 
 \displaystyle{ 
  \left[\psi(l),\,\psi^\dagger(l^\prime)\right]
  = \delta(l - l^\prime) \, , 
 }& \nonumber \\ 
 & 
 \displaystyle{ 
  \left[\psi(l),\,\psi(l^\prime)\right] = 0 
  = \left[\psi^\dagger(l),\,\psi^\dagger(l^\prime)\right] \, , 
 }& 
\end{eqnarray} 
and the ``vacuum" state $\left|0\right>$ that is annihilated by all $\psi(l)$ i.e., for which we have
$\psi(l) \left|0\right> = 0$ ($\left<0\right| \psi^\dagger(l) = 0$). 
 We then define the state $ \left|\Psi\right> $ by
\begin{eqnarray} 
 \left|\Psi\right> 
 &=& 
 \sum_{r=0}^\infty  
 \frac{1}{r !}
 \int_0^\infty dl_1\,\psi^\dagger(l_1) 
  \cdots \int_0^\infty dl_r\,\psi^\dagger(l_r)
 \left|0\right> 
 \left<w(l_1) \cdots w(l_r)\right> 
  \nonumber \\ 
 &=& 
 \exp 
 \left[ 
  \sum_{r=1}^\infty  
  \frac{1}{r !} 
  \int_0^\infty dl_1\,\psi^\dagger(l_1) 
   \cdots \int_0^\infty dl_r\,\psi^\dagger(l_r)\, 
  w^{(r)}\left( l_1, \cdots, l_r \right) 
 \right] 
 \left|0\right> \, , 
  \label{eq:def_Fvacuum}
\end{eqnarray} 
where $w^{(r)}\left( l_1, \cdots, l_r \right)$ 
represents the connected part of the Green function 
$\left< w(l_1) \cdots w(l_r) \right>$:\footnote{ 
 We recall that 
\begin{equation} 
 w^{(r)}(l_1,\,\cdots,l_r) 
 = {\cal O}\left((g_{\rm st})^{2(r-1)}\right) \, , 
  \label{eq:order_of_w_c}
\end{equation} 
in our convention. 
}
\begin{eqnarray} 
 && 
 \left<w(l)\right> = w^{(1)}(l) \, , 
  \nonumber \\ 
 && 
 \left<w(l_1) w(l_2)\right> 
 = 
 w^{(2)}(l_1,\,l_2) + w^{(1)}(l_1) w^{(1)}(l_2) \, , 
  \nonumber \\ 
 && \cdots \, . 
\end{eqnarray} 
 Using $\left|\Psi\right>$, 
the original Green functions can be expressed as 
\begin{equation} 
 \left< w(l_1) \cdots w(l_r) \right> 
 = 
 \left<0\right| \psi(l_1) \cdots \psi(l_r) \left|\Psi\right> \, , 
  \label{eq:corr_in_Psi}
\end{equation} 
and the continuum loop equations (\ref{eq:loop_equations_w}) can be written in the compact form 
\begin{equation} 
 D(l) \left|\Psi\right> = 0 \, , 
  \label{eq:loop_equation} 
\end{equation} 
where 
\begin{eqnarray} 
 && 
 D(l) \equiv l T(l) + \rho_s(l) \, , 
  \nonumber \\ 
 && 
 T(l) 
 \equiv 
 \int_0^l dl^\prime\,\psi(l^\prime)\,\psi(l - l^\prime) 
 + 
 g_{\rm st}^2 
 \int_0^\infty dl^\prime\,l^\prime\, 
  \psi^\dagger(l^\prime)\,\psi(l + l^\prime) \, . 
  \label{eq:loop_eq_csft} 
\end{eqnarray} 

 For subsequent considerations, 
we remark here 
that $T(l)$ is isomorphic to the energy-momentum tensor 
in momentum space 
if $l$ is regarded as a momentum. 
 To see this, 
we construct a new variable $\varphi(l)$ 
from $\psi(l)$ and $\psi^\dagger(l)$ as 
\begin{eqnarray} 
 & 
 \displaystyle{ 
  \varphi(l) 
  = 
  \left\{ 
   \begin{array}{ll} 
    \displaystyle{ 
     \frac{\sqrt{2}}{g_{\rm st}}\, \psi(l) 
    } & (l > 0) \\ 
     & \\ 
    \displaystyle{ 
     \frac{g_{\rm st}}{\sqrt{2}}\, (-l)\psi^\dagger(-l) 
    } & (l < 0) 
   \end{array} 
  \right. \, .
 }& 
\end{eqnarray} 
This variable satisfies the commutation relation 
\begin{equation} 
 \displaystyle{ 
  \left[ 
   \varphi(l),\,\varphi(l^\prime) 
  \right] = l \delta(l + l^\prime) \, . 
 } 
\end{equation} 
 $T(l)$ can then be rewritten as 
\begin{equation} 
 T(l) 
 = 
 g_{\rm st}^2 
 \int_{-\infty}^\infty dl^\prime\, 
  \frac{1}{2} :\varphi(l^\prime)\,\varphi(l - l^\prime): \, . 
\end{equation} 
 If we further introduce their ``coordinate'' representation by  
\begin{eqnarray} 
 \widetilde{\varphi}(\sigma) 
 &=& 
 \frac{1}{i} \int_{-\infty}^\infty \frac{dl}{l}\, 
  e^{i l \sigma}\,\varphi(l) \, , 
  \nonumber \\ 
 \widetilde{T}(\sigma) 
 &=& 
 \int_{-\infty}^\infty dl\, e^{i l \sigma}\, T(l) \, , 
\end{eqnarray} 
$\widetilde{T}(\sigma)$ can be expressed as 
\begin{equation} 
 \widetilde{T}(\sigma) 
 = 
 g_{\rm st}^2\, 
 \frac{1}{2} :(\del_\sigma \widetilde{\varphi}(\sigma))^2: 
  \, , 
\end{equation} 
and the two-point Green function 
of $\del_\sigma \widetilde{\varphi}(\sigma)$ is given by 
\begin{equation} 
 \left< 0 \right| 
  \del_\sigma \widetilde{\varphi}(\sigma) 
  \del_{\sigma^\prime} \widetilde{\varphi}(\sigma^\prime) 
 \left| 0 \right> 
 = 
 \frac{1}{\left(\sigma - \sigma^\prime\right)^2}\, . 
\end{equation} 
 Hence, $\widetilde{T}(\sigma)$ is the energy-momentum tensor 
of a free massless scalar field $\widetilde{\varphi}(\sigma)$ 
up to an overall factor of $g_{\rm st}^2$. 

\subsection{Classical approximation of the loop equations} 
\label{subsec:classical}

 This subsection deals with the classical limit, 
$g_{\rm st} \rightarrow 0$, of the loop equations. 
 In the perturbative expansion, the amplitude $w^{(1)}(l)$ of 
one external loop begins  
with the disk contribution $w^{(1)}_0(l)$, 
which is ${\cal O}((g_{\rm st})^0)$ in our convention 
(\ref{eq:order_of_w_c}). 
 Because $w^{(2)}(l_1,\,l_2) = {\cal O}(g_{\rm st}^2)$ 
we have
\begin{equation} 
 \left<w(l_1) w(l_2)\right> 
 = w^{(1)}_0(l_1) w^{(1)}_0(l_2) 
   + {\cal O}\left(g_{\rm st}^2\right) \, . 
\end{equation} 
 Thus, Eq.~(\ref{eq:loop_equations_w}) with $r = 0$ 
at ${\cal O}((g_{\rm st})^0)$ becomes 
\begin{equation} 
 \del_\zeta \left(\widetilde{w}(\zeta)\right)^2 
 + {\cal O}((g_{\rm st})^2) 
 = 
 3 \zeta^2 - \frac{3}{4}\,t \, , 
\end{equation} 
where $\widetilde{w}(\zeta)$ is the Laplace transform 
of $w^{(1)}(l)$. 
 This equation determines $\left(\widetilde{w}(\zeta)\right)^2$ 
up to a constant $\varpi$:
\begin{eqnarray} 
 \left(\widetilde{w}(\zeta)\right)^2 
 &=& 
 \left(\widetilde{w}_0(\zeta)\right)^2 + \varpi \, , 
  \nonumber \\ 
 \left(\widetilde{w}_0(\zeta)\right)^2 
 &=& 
 \left(\zeta - \frac{\sqrt{t}}{2}\right)^2 
 \left(\zeta + \sqrt{t}\right) \, .  
  \label{eq:loop_amplitude}
\end{eqnarray}  
 In the classical limit, 
the loop equations do not place a restriction on $\varpi$. 

 As we have seen in \S\ref{sec:inst}, 
in the matrix model, 
$\varpi$ is determined by the number of eigenvalues 
positioned on the top of the potential. 
 By contrast, 
in the classical approximation of the loop equations, 
$\varpi$ is ambiguous. 
 This is in some senses obvious from the beginning.
It is clear that $\varpi$ cannot be described in terms of classical 
solutions of closed string field, because it is not of order $g_{\rm st}^2$ 
but rather $g_{\rm st}$.
\subsection{Loop equations to all orders and the nonperturbative effect}
\label{subsec:beyond_classical}
 Beginning in  this subsection, 
we investigate 
the possibility of determining 
the constant $\varpi$ and the chemical potential 
using the loop equations to all orders,(\ref{eq:loop_equation}). 
 
 First, we examine the vacuum expectation value (VEV) 
of an operator ${\cal A}$ in the presence of a fixed number 
of instantons. 
 This means that we pick up, say, a single-instanton sector 
out of an infinite number of sectors 
and consider the VEV of ${\cal A}$ in this sector. 
 In the matrix model, such a quantity can be written as 
\begin{eqnarray} 
 \left< {\cal A} \right>^{({\rm 1\mbox{-}inst})} 
 &=& 
 \frac{Z_{N-1}^{\prime\quad ({\rm 0\mbox{-}inst})}} 
      {Z_N} 
 \times 
 N \int_{x \sim x_{(0)}} dx\,  
 \left< 
  {\cal A}(\phi^\prime,\,x)\,\det (x - \phi^\prime)^2 
 \right>^{\prime\,({\rm 0\mbox{-}inst})}\ e^{-NV(x)}
  \nonumber \\ 
 &\propto& 
 N \int_{x \sim x_{(0)}} dx\, 
 \left< 
  {\cal A}(\phi^\prime,\,x)\, 
  \exp 
  \left[ 
   2\,{\rm tr}\,{\rm ln} (x - \phi^\prime) 
  \right] 
 \right>^{\prime\,({\rm 0\mbox{-}inst})} \ e^{-NV(x)}
  \nonumber \\ 
 &=& 
 N \int_{x \sim x_{(0)}} dx\, \ e^{-NV(x)}
  \nonumber \\ 
 && \qquad 
 \times 
 \left< 
  {\cal A}(\phi^\prime,\,x)\, 
  \exp 
  \left[ 
   - 2\,\int_0^\infty dl\, 
   \left( 
    \frac{e^{-x l}}{l}\, {\rm tr}\,e^{l \phi^\prime} 
    - 
    \frac{e^{-l}}{l}\, {\rm tr}\,1 
   \right) 
  \right] 
 \right>^{\prime\,({\rm 0\mbox{-}inst})} \, , 
  \nonumber \\ 
  \label{eq:vev_matrix_model} 
\end{eqnarray} 
where $Z_{N-1}^{\prime\quad({\rm 0\mbox{-}inst})}$ 
and $\left<,\,\right>^{\prime\,({\rm 0\mbox{-}inst})}$ 
are defined in Eq.~(\ref{eq:average_0-inst}). 

 Because we instead analyze the loop equations here, 
we need to construct a state describing the single-instanton vacuum 
by closely considering the expression (\ref{eq:vev_matrix_model}). 
 Equation (\ref{eq:vev_matrix_model}) 
expresses 
the VEV of an operator ${\cal A}$ in the single-instanton vacuum 
in terms of 
the VEV in the null instanton vacuum of the operator 
obtained by making the loop of length $l$ (${\rm tr}\,e^{l \phi^\prime}$) 
condensed with a weight $-\frac{2\,e^{-x l}}{l}$ 
and integrating with respect to $x$ around $x_{(0)}$. 
 In the continuum limit, 
the null instanton vacuum state 
$\left|\Psi\right>_{\rm 0\mbox{-}inst}$ is given by 
\begin{equation} 
 \left|\Psi\right>_{\rm 0\mbox{-}inst} 
 = 
 \exp 
 \left[ 
  \sum_{r=1}^\infty  
  \frac{1}{r !} 
  \int_0^\infty dl_1\,\psi^\dagger(l_1) 
   \cdots \int_0^\infty dl_r\,\psi^\dagger(l_r)\, 
  \left. w^{(r)}\left( l_1, \cdots, l_r \right) \right|_{\rm pert} 
 \right] 
 \left|0\right> \, , 
\end{equation} 
where $\left.w^{(r)}(l_1,\cdots,\,l_r)\right|_{\rm pert}$ 
is $w^{(r)}(l_1,\cdots,\,l_r)$ in the null instanton vacuum. 
 The single-instanton vacuum state can be obtained by 
applying the operator ${\cal V}(\zeta)$ 
to $\left|\Psi\right>_{\rm 0\mbox{-}inst}$,
which condenses the loops with an appropriate weight, 
and then integrating 
${\cal V}(\zeta)\left|\Psi\right>_{\rm 0\mbox{-}inst}$ 
with respect to $\zeta$ 
around $\frac{\sqrt{t}}{2}$,corresponding to $x_{(0)}$. 
 The operator ${\cal V}(\zeta)$ 
is obtained by replacing 
the operator $\frac{1}{N}\,{\rm tr}\,e^{l \phi^\prime}$ 
in the exponent of Eq.~(\ref{eq:vev_matrix_model}) 
with $\psi(l)$: 
\begin{equation} 
 {\cal V}(\zeta) 
 = 
 C_{\rm inst} 
 \exp 
 \left[ 
  - Q \int_0^\infty dl\,\frac{2}{l}\,e^{-\zeta l}\,\psi(l) 
 \right] 
  \, . 
  \label{eq:operatorV} 
\end{equation} 
Here, we have introduced the parameter $Q$, 
to take account of a possible renormalization 
of the order $N \sim \frac{1}{g_{\rm st}}$, 
and the overall normalization $C_{\rm inst}$. 
 Using ${\cal V}(\zeta)$ in Eq.~(\ref{eq:operatorV}), 
we obtain the single-instanton vacuum in the form 
\begin{eqnarray} 
 \left|\Psi\right>_{\rm 1\mbox{-}inst} 
 &=& 
 \int_{\zeta \sim \frac{\sqrt{t}}{2}} d\zeta\, 
  {\cal V}(\zeta) \left|\Psi\right>_{\rm 0\mbox{-}inst} 
  \nonumber \\ 
 &=& 
 C_{\rm inst} 
 \int_{\zeta \sim \frac{\sqrt{t}}{2}} d\zeta\, 
 \exp 
 \left[ 
  - Q \int_0^\infty dl\, \frac{2}{l}\,e^{-\zeta l}\, \psi(l) 
 \right] 
 \left|\Psi\right>_{\rm 0\mbox{-}inst} \, .  
  \label{eq:inst_op_naive} 
\end{eqnarray} 

 Our investigation to this point has not considered
the all-order loop equations (\ref{eq:loop_equation}). 
 We explain in the following subsections 
that $Q$ is identical to
the constant ambiguity $\varpi$ in the loop amplitude
and that $C_{\rm inst}$ is related to the chemical potential. 
 In the rest of this subsection, 
we would like to find 
whether $Q$ and $C_{\rm inst}$ 
can be determined by analyzing the all-order loop equations. 

 As we have seen above, 
the operator ${\cal V}(\zeta)$ in Eq.~(\ref{eq:operatorV}) 
has been proposed to introduce an additional instanton into 
the system if $\zeta$ is integrated around 
$\frac{\sqrt{t}}{2}$. 
 From the viewpoint of the matrix model, 
$ 
\displaystyle{ 
 \int_{\zeta \sim \frac{\sqrt{t}}{2}} d\zeta\,{\cal V}(\zeta) 
}$ 
plays the role of adding an eigenvalue 
at the top of the potential. 
 Hence, 
$ 
\displaystyle{ 
 \int_{-\infty}^\infty d\zeta\,{\cal V}(\zeta) 
}$ 
adds an eigenvalue without specifying its position. 
 We recall that 
the full vacuum $\left|\Psi\right>$ consists 
of the vacua of various numbers of instantons: 
\begin{equation} 
 \left|\Psi\right> = 
 \left|\Psi\right>_{\rm 0\mbox{-}inst} 
 + 
 \left|\Psi\right>_{\rm 1\mbox{-}inst} 
 + 
 \left|\Psi\right>_{\rm 2\mbox{-}inst} 
 + \cdots \, . 
\end{equation} 
 Thus, we should have 
\begin{equation} 
 \left|\Psi\right>^{(N+1)} 
 = 
 \int_{-\infty}^\infty d\zeta\,{\cal V}(\zeta) 
 \left|\Psi\right>^{(N)} \, , 
\label{eqn:plusone} 
\end{equation} 
where $\left|\Psi\right>^{(N)}$ is the vacuum 
of the system consisting of $N$ eigenvalues. 
 In the large-$N$ limit, 
this equation requires 
that if $\left|\Psi\right>$ is a solution 
of the loop equations (\ref{eq:loop_equation}), 
$\displaystyle{ 
 \int_{-\infty}^\infty d\zeta\,{\cal V}(\zeta) \left|\Psi\right> 
}$ must also be a solution of these equations. 
 This will be the case if 
\begin{eqnarray} 
 \left[\widetilde{D}(\zeta),\,{\cal V}(\zeta)\right] 
 \sim \del_\zeta {\cal V}(\zeta) \, 
\end{eqnarray} 
holds, where $\widetilde{D}(\zeta)$ is the Laplace transform of $D(l)$ .
 As remarked in \S\ref{subsec:loop_eq}, 
the operator part of $\widetilde{D}(\zeta)$ is 
essentially 
the energy-momentum tensor $\widetilde{T}(\zeta)$. 
 Thus, the above relation implies that 
${\cal V}(\zeta)$ 
should be a primary field of conformal dimension equal to $1$. 
 The works in Ref.~\cite{Fukuma:1996hj} 
carry out a search for such an operator ${\cal V}(\zeta)$ 
and find an expression in terms of local operators. 
 We find 
that the primary operator with conformal dimension 1 
that resembles 
Eq.~(\ref{eq:inst_op_naive}) is 
\begin{equation} 
 :\exp 
  \left(-i \sqrt{2}\,\widetilde{\varphi}(\sigma) \right): 
 \,=\, 
 :\exp 
  \left[ 
   - 
   \int_0^\infty dl\, 
   \left( 
    \frac{2}{g_{\rm st} l}\, e^{i l \sigma}\,\psi(l) 
    - 
    g_{\rm st}\,e^{-i l \sigma}\, \psi^\dagger(l) 
   \right) 
  \right]: \, . 
\end{equation} 
 To bring this into the form of Eq.~(\ref{eq:inst_op_naive}), 
we apply the analytic continuation $\sigma \rightarrow i \zeta$,
which yields 
\begin{equation} 
 {\cal V}(\zeta) 
 \equiv 
 C_{\rm inst} 
 :\exp 
  \left[ 
   - 
   \int_0^\infty dl\, 
   \left( 
    \frac{2}{g_{\rm st} l}\, e^{- \zeta l}\,\psi(l) 
    - 
    g_{\rm st}\,e^{\zeta l}\, \psi^\dagger(l) 
   \right) 
  \right]: \, . 
  \label{eq:operator_V}
\end{equation} 
 We thus find that 
the single-instanton vacuum 
$\left|\Psi\right>_{\rm 1\mbox{-}inst}$ is given by 
\begin{eqnarray} 
 && 
 \left|\Psi\right>_{\rm 1\mbox{-}inst} 
 = 
 D_{\rm inst} \left|\Psi\right>_{\rm 0\mbox{-}inst}, 
  \nonumber \\ 
 && 
 D_{\rm inst} 
 = 
 C_{\rm inst} 
 \int_{\zeta \sim \frac{\sqrt{t}}{2}} d\zeta 
 :\exp 
  \left[ 
   \int_0^\infty dl 
    \left( 
     - \frac{2}{g_{\rm st} l}\,e^{-\zeta l}\,\psi(l) 
     + g_{\rm st}\,e^{\zeta l}\,\psi^\dagger(l) 
    \right)  
  \right]: \, . 
  \label{eq:state_for_instanton}
\end{eqnarray}  
 The expression (\ref{eq:state_for_instanton}) shows that 
the all-order loop equations determine 
$Q$ ($Q = \frac{1}{g_{\rm st}}$) 
appearing in Eq.~(\ref{eq:inst_op_naive}). 
 However, 
the state 
$\displaystyle{ 
 \int_{-\infty}^\infty d\zeta\,{\cal V}(\zeta) 
 \left|\Psi\right>  
}$ 
with ${\cal V}(\zeta)$ in Eq.~(\ref{eq:operator_V}) 
is always a solution of the loop equations 
for any value of $C_{\rm inst}$. 
 Thus, the normalization constant $C_{\rm inst}$ 
cannot be determined 
as long as we use only closed string fields.

\subsection{Loop amplitude in closed string field theory} 
\label{subsec:loop_amp}
 Now we know how to express the instanton contributions in the loop  
equation formalism, at least up to an overall normalization. 
 In this subsection, we examine how they appear 
in the limit $g_{\rm st}\rightarrow 0$. 
 Then, we compare the results of the loop equation with those of the 
matrix model and the Liouville theory approaches. 
 In particular, 
as we discussed in \S\ref{subsec:classical}, they should correspond 
to the solution of Eq.~(\ref{eq:loop_amplitude}) with some $\varpi$. 
 We calculate the value of $\varpi$ using the loop equation and 
compare it with that obtained in the other approaches. 

 In order to study the limit $g_{\rm st}\rightarrow 0$, 
it is convenient to express 
the state (\ref{eq:state_for_instanton}) 
using Eq.~(\ref{eq:def_Fvacuum}) as follows:
\begin{eqnarray} 
 \left|\Psi\right>_{\rm 1\mbox{-}inst} 
 &=& 
 C_{\rm inst} 
 \int_{\zeta\sim \frac{\sqrt{t}}{2}} d\zeta\, 
 \exp 
 \left( 
  \int_0^\infty dl\, 
   g_{\rm st}\, e^{\zeta l}\, \psi^\dagger(l) 
 \right) 
  \nonumber \\ 
 && \qquad 
 \times 
 \exp 
 \left[ 
  \sum_{r=1}^\infty 
  \frac{1}{r !} 
  \int_0^\infty d l_1\, 
   \left( 
    \psi^\dagger(l_1) - \frac{2}{g_{\rm st} l_1}\,e^{-\zeta l_1} 
   \right) 
   \times \cdots 
 \right. 
  \nonumber \\ 
 && \qquad \qquad \qquad \quad 
 \left. 
  \times 
  \int_0^\infty d l_r\, 
   \left( 
    \psi^\dagger(l_r) - \frac{2}{g_{\rm st} l_r}\,e^{-\zeta l_r} 
   \right) 
 \right. 
  \nonumber \\ 
 && \qquad \qquad \qquad \qquad  
 \left. 
  \times 
  \left. w^{(r)}(l_1, \cdots, l_r) \right|_{\rm pert} 
 \right] 
 \left|0\right>\, . 
  \label{eq:Psi_inst_another_expr}
\end{eqnarray} 
 Because $w^{(r)}(l_1,\,\cdots,l_r) 
= {\cal O}\left((g_{\rm st})^{2(r-1)}\right) $, 
we can expand Eq.(\ref{eq:Psi_inst_another_expr}) in terms of 
$g_{\rm st}$. 
 Using this expression, let us calculate 
$\left< 0 | \Psi \right>_{\rm 1\mbox{-} inst}$ and 
$\left< 0 \right| \psi(l) \left|\Psi\right>_{1\mbox{-}\rm inst}$, 
which can be compared with the quantities calculated in 
the matrix model formulation. 

The quantity $\left< 0 | \Psi \right>_{\rm 1\mbox{-} inst}$ 
should be the continuum version of  
$\int_{x\sim x_{(0)}} dx\, e^{-V_{\rm eff}(x)}$ 
in the matrix model, 
and it can be expressed as 
$\int_{\zeta\sim \frac{\sqrt{t}}{2}} d\zeta\, e^{-V_{\rm eff}(\zeta )}$ 
in terms of the continuum effective potential $V_{\rm eff}(\zeta )$. 
 Thus, using Eq.~(\ref{eq:Psi_inst_another_expr}), we find that 
$V_{\rm eff}(\zeta )$ should be expanded as
\begin{eqnarray} 
 - V_{\rm eff}(\zeta) 
 &=& 
 \int_0^\infty dl\,\left. w^{(1)}(l) \right|_{\rm pert}\, 
 \left( -\frac{2}{g_{\rm st} l}\,e^{-\zeta l} \right) 
  \nonumber \\ 
 && \ 
 + 
 \frac{1}{2}\, 
 \int_0^\infty d l_1\,\frac{2}{g_{\rm st} l_1}\,e^{-\zeta l_1}\, 
 \int_0^\infty d l_2\,\frac{2}{g_{\rm st} l_2}\,e^{-\zeta l_2}\, 
  \left. w^{(2)}(l_1,\,l_2) \right|_{\rm pert} 
  \nonumber \\ 
 && \ 
 + \cdots \, . 
\label{veff from loop}
\end{eqnarray} 
 The leading-order contribution to $w^{(1)}(l)$ 
is from the disk amplitude $w^{(1)}_0(l)$. 
 Thus, the above formula to  leading order coincides with 
the matrix model result in the limit $N\rightarrow \infty$. 
 The leading-order contribution to the second term on the right-hand side 
is from the cylinder amplitude, $w^{(2)}_0(l_1,\,l_2)$. 
 Higher-order terms involve 
the contributions from worldsheets with more boundaries. 
 Therefore, the expansion above corresponds to the expansion 
 Eq.~(\ref{eqn:qeffect})
in the matrix model. 

The quantity $\left< 0 \right| \psi(l) \left|\Psi\right>_{1\mbox{-}\rm inst}$ 
should correspond to the loop amplitude in the instanton background 
$\left. w(l) \right|_{\rm inst}$. 
 It is calculated as 
\begin{eqnarray} 
\left. w(l) \right|_{\rm inst} 
 &=& 
 \left< 0 \right| \psi(l) \left|\Psi\right>_{1\mbox{-}\rm inst} 
  \nonumber \\ 
 &=& 
 C_{\rm inst} \int_{\zeta\sim \frac{\sqrt{t}}{2}} d\zeta\, 
 e^{- V_{\rm eff}(\zeta)} \, 
  \nonumber \\ 
 && \quad 
 \times 
 \biggl[ 
  g_{\rm st} e^{\zeta\,l} 
    \nonumber \\ 
 && \qquad 
  + 
  \sum_{r=1}^\infty 
  \frac{1}{(r-1)!}\,
   \int_0^\infty dl_1 
    \left(-\frac{2}{g_{\rm st} l_1}\,e^{-\zeta l_1}\right) 
   \cdots 
   \int_0^\infty dl_{r-1} 
    \left(-\frac{2}{g_{\rm st}l_{r-1}}\,e^{-\zeta l_{r-1}}\right) 
  \nonumber \\ 
 && \qquad \qquad \qquad \qquad \qquad \qquad  
  \times 
  \left. w^{(r)}(l_1,\cdots,l_{r-1},\,l) \right|_{\rm pert} 
 \biggl] 
 \, . 
  \label{eq:loop_amp_sft}
\end{eqnarray} 
 The first few terms in the $g_{\rm st}$ expansion are given as 
\begin{eqnarray}
\left. w(l) \right|_{\rm inst} 
 & = & C_{\rm inst} \int_{\zeta\sim \frac{\sqrt{t}}{2}} d\zeta\, 
       e^{- V_{\rm eff}(\zeta)} \, \nonumber \\
 & \times & \left(\left. w^{(1)}(l) \right|_{\rm pert} 
           +g_{\rm st} e^{\zeta\,l}
             +\int_0^\infty dl_1 
              \left(-\frac{2}{g_{\rm st} l_1}\,e^{-\zeta l_1}\right) 
              \left. w^{(2)}_0(l_1,\,l) \right|_{\rm pert}      
            +\cdots
            \right). \nonumber \\          
\end{eqnarray} 
 In the limit $g_{\rm st}\rightarrow 0$, we can use the saddle point 
approximation to evaluate the integration over $\zeta$. 
 The saddle point $\zeta =\sqrt{t}/2$ can be identified 
with $x_{(0)}$ in the matrix model. 
 Thus, by Laplace transforming the above expression, 
it is easy to see that the three terms in it coincide with the 
matrix model results Eqs.~(\ref{eqn:0-instRin1-instR}) and (\ref{eqn:DeltaR}). 

 From the analysis to this point, it is obvious that what we have been doing 
here is exactly the continuum version of what we did in \S\ref{sec:inst}. 
 To be more concrete, Eq.~(\ref{eqn:plusone}) should be considered 
 as the continuum 
version of the manipulation used to isolate the contribution of one eigenvalue 
by introducing $q_i$ and $\bar{q}_i$. 
 The value $\zeta$ in Eq.~(\ref{eqn:plusone}) corresponds to the eigenvalue 
 $x$ in the continuum limit. 

 The cylinder amplitude $w^{(2)}_0(l_1,\,l_2)$ can be obtained by solving  
Eq.~(\ref{eq:loop_equations_w}). It is given as  
\begin{equation} 
 \widetilde{w}^{(2)}_0(\zeta_1,\,\zeta_2) 
 = 
 - g_{\rm st}^2\, 
  \del_{\zeta_1} \del_{\zeta_2}\, 
 {\rm ln} 
 \left( 
  \sqrt{\zeta_1 + \sqrt{t}} + \sqrt{\zeta_2 + \sqrt{t}} 
 \right) \, 
  \label{eq:cylinder_amp}
\end{equation} 
in the Laplace transformed form. 
 Thus 
$\widetilde{w}|_{\rm inst}(\zeta )=
\int_0^\infty dl e^{-\zeta l}w|_{\rm inst}(l)$ is obtained as 
\begin{eqnarray} 
 && 
 \widetilde{w}_0(\zeta) 
 + 
 g_{\rm st} 
 \left( 
  \frac{1}{\zeta - \frac{\sqrt{t}}{2}} 
  - 
  2\, \del_\zeta 
  {\rm ln}\, 
  \left( 
   \sqrt{\zeta + \sqrt{t}} + \sqrt{\frac{3}{2}}\,t^{\frac{1}{4}} 
  \right) 
 \right) 
 + {\cal O}((g_{\rm st})^2) 
  \nonumber \\ 
 && \quad 
 = 
 \widetilde{w}_0(\zeta) 
 + 
 \frac{1}{2}\,  
 \frac{g_{\rm st} \sqrt{6}\,t^{\frac{1}{4}}} 
      {\widetilde{w}_0(\zeta)} 
 + {\cal O}((g_{\rm st})^2) \, . 
  \label{eq:g_s_corr_in_loop_eq}
\end{eqnarray} 
 By comparing this with the expansion of $\widetilde{w}(\zeta)$ 
in $\varpi$ given by Eq.~(\ref{eq:loop_amplitude}), 
\begin{equation} 
 \widetilde{w}(\zeta) 
 = 
 \widetilde{w}_0(\zeta) 
 + 
 \frac{1}{2}\, \frac{\varpi}{\widetilde{w}_0(\zeta)} 
 + 
 {\cal O}((g_{\rm st})^2) \, , 
\end{equation} 
we find that 
\begin{equation} 
 \left. \varpi \right|_{\rm loop\ equation} 
 = g_{\rm st} \sqrt{6}\, t^{\frac{1}{4}}\, . 
  \label{eq:varpi_via_loop_eqs}
\end{equation} 
 Taking the factor in Eq.~(\ref{eq:g_S_translate}) 
into account, 
this value of $\varpi$ 
is found to coincide with the matrix model result 
(\ref{eqn:DeltaRvalue}): 
\begin{equation} 
 \left. \varpi \right|_{\rm matrix} 
 = g_s \frac{3 \sqrt{3}}{4}\,t^{\frac{1}{4}} \, . 
  \label{eq:b_in_matrix_model}
\end{equation} 

 This value of $\varpi $ is consistent with the result of the Liouville 
theory. 
 In the Liouville theory \cite{Knizhnik:ak}, the instanton we have been 
studying corresponds to the D-instanton. 
 The amplitudes in the presence of such a D-brane can be calculated by 
using the open string theory. 
 In particular, $w(l)|_{\rm inst}$ can be evaluated as an expansion 
with respect to the string coupling constant. 
 The first correction to the disk amplitude is given by the cylinder amplitude 
with one boundary on the D-brane. 
 Thus $\varpi$ can be obtained by calculating such a cylinder amplitude. 
 This is exactly what we have done above. 

 In the Liouville theory \cite{Knizhnik:ak}, 
the D-instanton corresponds to the ZZ-brane boundary state 
$\left|B_{(m,\,n)}\right>_{\rm ZZ}$ with $(m,\,n) = (1,\,1)$. 
 The loop $w(l)$ that we have been discussing corresponds to the 
FZZT boundary state $\left|B_s\right>_{\rm FZZT}$. 
 The normalization of $\left|B_{(m,\,n)}\right>_{\rm ZZ}$ 
and $\left|B_s\right>_{\rm FZZT}$ 
can be fixed through a  modular bootstrap or, in other words, 
the open-closed duality \cite{Zamolodchikov:2001ah}. 
 Therefore, we need to 
calculate the cylinder amplitude with one boundary on the ZZ-brane and 
the other on the FZZT-brane. 
 This can be calculated in a manner similar to that employed in Ref.\cite{Martinec:2003ka}, 
but here we proceed differently, as we now demonstrate. 
 It is known that the properly normalized boundary states 
$\left|B_{(m,\,n)}\right>_{\rm ZZ}$ and 
$\left|B_s\right>_{\rm FZZT}$ satisfy the following relation
\cite{Martinec:2003ka,Klebanov:2003wg,Seiberg:2003nm} : 
\begin{equation} 
 \left|B_{(m,\,n)}\right>_{\rm ZZ} 
 = 
 \left|B_{s=i\left(\frac{m}{b} + nb\right)}\right>_{\rm FZZT} 
 - 
 \left|B_{s=i\left(\frac{m}{b} - nb\right)}\right>_{\rm FZZT} 
 \, . 
  \label{eq:rel_ZZ_FZZT} 
\end{equation}  
 We have $b = \sqrt{\frac{2}{3}}$ for the $c=0$ case, 
and $s$ is related to the boundary cosmological constant $\mu_B$ 
and the bulk cosmological constant $\mu$ 
in the Liouville theory as 
\begin{equation} 
 \cosh(\pi b s) 
 = 
 \frac{\mu_B}{\sqrt{\mu}}\, 
 \left(\sin(\pi b^2)\right)^{\frac{1}{2}}\, . 
  \label{eq:s_mu_B} 
\end{equation} 
 The relation (\ref{eq:rel_ZZ_FZZT}) can be used to 
calculate the cylinder amplitude in question 
from $w^{(2)}_0(l_1,\,l_2)$ obtained using the loop equation. 
 Thus, we can check if our results are consistent with 
those obtained from the Liouville approach. 

 The value of $\mu_B$, or equivalently $s$, is obviously related to the 
variable $\zeta$ we have been using. 
 Because $b = \sqrt{\frac{2}{3}}$,  
$s = i\left(\frac{1}{b} \pm b\right)$ gives 
$\cosh(\pi b s)=\frac{1}{2}$. 
 On the other hand, 
the Laplace transformation of the disk amplitude $\widetilde{w}_0$ 
is proportional to $\cosh(\pi s/b)$. 
 Thus we see that $\widetilde{w}_0$ vanishes if 
$s = i\left(\frac{1}{b} \pm b\right)$. 
 Therefore $s = i\left(\frac{1}{b} \pm b\right)$ should correspond to 
$\zeta = \frac{\sqrt{t}}{2}$. 
 A careful analysis shows that 
$s = i \left(\frac{1}{b} + b\right)$ 
corresponds to $\zeta = \frac{\sqrt{t}}{2}$ on the second Riemann sheet, 
while 
$s = i \left(\frac{1}{b} - b\right)$ corresponds to that 
on the first Riemann sheet. 

 Now we calculate the cylinder amplitude. 
 The FZZT-brane boundary state can be naturally identified with the 
boundary formed by 
$-\tr\ln(z-\phi)$ in the double-scaled matrix model 
including the normalization. 
 Therefore, the ${\cal O}(g_{\rm st})$ term in $\widetilde{w}(\zeta)$ 
should be 
\begin{equation} 
 \int_0^\infty dl\,e^{-\zeta l} 
 \left[ 
  \int_0^\infty \frac{dl^\prime}{l^\prime} 
   \,e^{-\zeta^\prime l^\prime} 
   \frac{w^{(2)}_0(l,\,l^\prime)}{g_{\rm st}} 
 \right]_{\zeta^\prime 
          = \frac{\sqrt{t}}{2}\ 
            {\rm on\ the\ first\ Riemann\ sheet} 
         }^{\zeta^\prime 
            = \frac{\sqrt{t}}{2}\ 
            {\rm on\ the\ second\ Riemann\ sheet}} \, .  
            \label{eq:order_g_st}
\end{equation} 
 Substituting the cylinder amplitude given by Eq.~(\ref{eq:cylinder_amp}) into Eq.~(\ref{eq:order_g_st}), 
we obtain 
\begin{eqnarray} 
 && 
 \left[ 
  \frac{g_{\rm st}}{2}\, 
  \frac{\frac{1}{\sqrt{\zeta + \sqrt{t}}}} 
       {\sqrt{\zeta + \sqrt{t}} + \sqrt{\zeta^\prime + \sqrt{t}}} 
 \right]_{\zeta^\prime 
          = \frac{\sqrt{t}}{2}\ 
            {\rm on\ the\ first\ Riemann\ sheet} 
         }^{\zeta^\prime 
            = \frac{\sqrt{t}}{2}\ 
            {\rm on\ the\ second\ Riemann\ sheet}} 
  \nonumber \\ 
 && \quad 
 = 
 \frac{g_{\rm st}}{2}\, 
 \frac{1}{\sqrt{\zeta + \sqrt{t}}} 
 \left( 
  \frac{1}{\sqrt{\zeta + \sqrt{t}} - \sqrt{\frac{3}{2}\,\sqrt{t}}} 
  - 
  \frac{1}{\sqrt{\zeta + \sqrt{t}} + \sqrt{\frac{3}{2}\,\sqrt{t}}} 
 \right) 
  \nonumber \\ 
 && \quad 
 = 
 \frac{1}{2}\, 
 \frac{g_{\rm st} \sqrt{6}\,t^{\frac{1}{4}}} 
      {\widetilde{w}_0(\zeta)} \, . 
\end{eqnarray} 
 This shows that 
$\varpi = g_{\rm st}\sqrt{6}\,t^{\frac{1}{4}}$, as anticipated. 

 The relation in 
Eq.~(\ref{eq:rel_ZZ_FZZT}) can be seen more directly 
from the matrix model point of view. 
 Specifically, as we have seen in Eq.~(\ref{eq:vev_matrix_model}), 
adding a single boundary corresponding to the ZZ-brane 
amounts to inserting $2\,{\rm Re}\,\tr\ln(z-\phi)$ 
evaluated at $z = x_{(0)}$ 
from the matrix model point of view: 
\begin{equation}
 \left|B_{(1,\,1)}\right>_{\rm ZZ} 
 \longleftrightarrow 
 \left. 
  2\,{\rm Re}\,\tr\ln(z-\phi) 
 \right|_{z=x_{(0)}}.
\label{eqn:MMZZ} 
\end{equation} 
 On the other hand, as we noted above, 
the FZZT brane 
corresponds to the integral of the resolvent from 
the first sheet to the second sheet: 
\begin{eqnarray}
 \left|B_{s=i\left(\frac{1}{b} + b\right)}\right>_{\rm FZZT} 
 - 
 \left|B_{s=i\left(\frac{1}{b} - b\right)}\right>_{\rm FZZT} 
 & \longleftrightarrow & 
-\int_{z=x_{(0)}\ {\rm on\ the\ first\ sheet}}
     ^{z=x_{(0)}\ {\rm on\ the\ second\ sheet}}       
      dz\,\tr\frac{1}{z-\phi} \nonumber \\
 & = & 
 \left. 
  2\,{\rm Re}\,\tr\ln(z-\phi)
 \right|_{z=x_{(0)}}.
\label{eqn:MMFZZT}       
\end{eqnarray}
 Because the relation (\ref{eq:rel_ZZ_FZZT}) 
was obtained by using the open-closed duality, 
Eqs.~(\ref{eqn:MMZZ}) and (\ref{eqn:MMFZZT}) suggest 
that the information concerning the open-closed duality is somehow incorporated 
in the matrix model. 
 
\subsection{Ambiguity in the normalization of single-instanton vacuum state} 
\label{subsec:cp_loop_eq} 
 
 As we have seen in \S\ref{subsec:beyond_classical}, 
we cannot determine the constant $C_{\rm inst}$ using even the 
all-order loop equation. 
 For this reason, we cannot determine the chemical potential $\mu$ 
using this approach. 
 However, as we have shown in \S\ref{sec:chemical_pot}, $\mu$ can be 
determined as a universal quantity in the matrix model approach. 
 Therefore, if in the loop equation approach 
 we can mimic as closely as possible the procedure applied
 in the case of matrix model, we should be 
able to calculate $\mu$. 
 As we argued in the previous subsection, Eq.~(\ref{eqn:plusone}) 
 can be considered 
the continuum version of the manipulation used to pick up the contribution of 
one eigenvalue in the matrix model. 
 Hence, we can use this fact as a guide to obtain the continuum version of 
the calculation of the chemical potential. 
 Although $N$ is $\infty$ in the double scaling limit, here we consider 
$N$ to be large but finite and take the limit $N\rightarrow\infty$ later. 
 Dividing the integration region into two parts as 
\begin{equation} 
 \left|\Psi\right>^{(N)} 
 = 
 \left(
 \int_{-\sqrt{t}}^\infty d\zeta\,{\cal V}(\zeta) 
 +\int_{-\infty}^{-\sqrt{t}} d\zeta\,{\cal V}(\zeta)
\right)
 \left|\Psi\right>^{(N-1)} \,  
\end{equation} 
in Eq.~(\ref{eqn:plusone})
(here we have replaced $N$ by $N-1$ for consistency with the matrix 
model formulation), 
we can regard the second term on the right-hand side as corresponding to the 
zero-instanton sector and the first term to the single-instanton sector. 
 In this formulation, $\mu$ can be derived from the ratio of the first term 
to the second term, and it does not depend on the overall normalization 
of the operator ${\cal V}$. 
 Let us examine whether we can calculate $\mu$ in this formulation. 

 In order to calculate $\mu$, we should evaluate 
$\left< 0 | \Psi \right>^{(N)}$, which can be expressed as 
$\int_{-\infty}^\infty d\zeta \, e^{-V_{\rm eff}(\zeta )}$. 
It is very easy to obtain identities similar to 
Eqs.~(\ref{eq:Psi_inst_another_expr}) and (\ref{veff from loop}) in this case. 
 Because $\zeta> -\sqrt{t}$ corresponds to the single-instanton sector and 
$-\sqrt{t}>\zeta$ does to the zero-instanton sector, we calculate  
$V_{\rm eff}(\zeta )$ separately for these two cases. 

 For $\zeta> -\sqrt{t}$, we can use Eq.~(\ref{veff from loop}) without 
any change. 
  In the limit $g_{\rm st}\rightarrow 0$, the leading contribution is 
from the disk amplitude, and it reproduces the matrix model result in the 
$N\rightarrow \infty$ limit, up to an additive constant. 
The cylinder contribution to $-V_{\rm eff}(\zeta)$ is obtained 
as 
\begin{equation}
-2{\rm ln} 
 \left( 
  2\sqrt{\zeta + \sqrt{t}} 
 \right) 
 .
\label{cylinder in Sec 4.5}
\end{equation}
 We expect that the cylinder contribution, which corresponds to the one-loop 
amplitude for an open string, should contribute to the chemical potential 
$\mu$. 

 For $-\sqrt{t}>\zeta$, in evaluating $\left< 0 | \Psi \right>^{(N)}$, 
we should fix the value of the loop amplitudes on the cut in the complex 
$\zeta$ plane. 
 For our purposes, here we should mimic the manner in which we dealt with such an ambiguity in the matrix model. 
 Doing this, we obtain 
\begin{eqnarray} 
 - V_{\rm eff}(\zeta) 
 &=& 
 \mbox{Re}_{\zeta}
 \int_0^\infty dl\,\left. w^{(1)}_0(l) \right|_{\rm pert}\, 
 \left( -\frac{2}{g_{\rm st} l}\,e^{-\zeta l} \right) 
  \nonumber \\ 
 && \ 
 + 
 \frac{1}{2}\, 
 \mbox{Re}_{\zeta_1}
 \int_0^\infty d l_1\,\frac{2}{g_{\rm st} l_1}\,e^{-\zeta_1 l_1}\, 
 \mbox{Re}_{\zeta_2}
 \int_0^\infty d l_2\,\frac{2}{g_{\rm st} l_2}\,e^{-\zeta_1 l_2}\, 
  \left. w^{(2)}_0(l_1,\,l_2) \right|_{\zeta_1=\zeta_2=\zeta} 
  \nonumber \\ 
 && \ 
 + \cdots \, . 
\label{veff in section 4.5}
\end{eqnarray} 
 Here, $\mbox{Re}_{\zeta}$ denotes the operation of taking the real 
part as a function of $\zeta$. 
 What is relevant for us is the difference $V_{\rm eff}(\frac{\sqrt{t}}{2}) 
-V_{\rm eff}(-\sqrt{t})$ and that the leading-order contribution precisely 
reproduces the matrix model result. 
 The second term can be evaluated as 
\begin{eqnarray}
&-\frac{1}{2}&
\left[
{\rm ln} 
 \left( 
  \sqrt{\zeta_1 + \sqrt{t}} + \sqrt{\zeta_2 + \sqrt{t}} 
 \right)
+
{\rm ln} 
 \left( 
  \sqrt{\zeta_1 + \sqrt{t}} - \sqrt{\zeta_2 + \sqrt{t}} 
 \right)
\right. 
\nonumber
\\
& &
\left.\left.
+{\rm ln} 
 \left( 
  -\sqrt{\zeta_1 + \sqrt{t}} + \sqrt{\zeta_2 + \sqrt{t}} 
 \right)
+
{\rm ln} 
 \left( 
  -\sqrt{\zeta_1 + \sqrt{t}} - \sqrt{\zeta_2 + \sqrt{t}} 
 \right)
\right]
 \right|_{\zeta_1=\zeta_2=\zeta}
\nonumber
\\
& &
\hspace{1cm}
=
-{\rm ln}
\left(
\zeta_1-\zeta_2
\right)
|_{\zeta_1=\zeta_2=\zeta},
\end{eqnarray}
which is divergent. 
 The cylinder contribution is, in a sense, continuous at $\zeta =-\sqrt{t}$, 
because Eq.~(\ref{cylinder in Sec 4.5}) is also divergent when 
$\zeta\rightarrow -\sqrt{t}$. 
 In any case, because of this divergence, we cannot reproduce the value of the 
chemical potential $\mu$ in this approach. 

 If we proceed ignoring this divergence, 
we obtain the chemical potential as 
\begin{equation}
\mu 
\sim 
t^{-\frac{1}{2}-\frac{1}{8}}
\times
N
\times
\infty^{-1},
\end{equation}
where the factor $t^{-\frac{1}{2}}$ comes from the cylinder contribution 
in Eq.~(\ref{cylinder in Sec 4.5}), $t^{-\frac{1}{8}}$ from the Gaussian 
integration around the saddle point, and $N$ from the combinatorial factor. 
 Therefore, the loop equation approach reproduces the essential part of 
$\mu$, i.e. $t^{-\frac{5}{8}}$, but it fails to reproduce the precise numerical 
factor. 
 What we have seen in this subsection is the continuum version of what we 
observed in the first part of \S\ref{sec:chemical_pot}. 
 In the matrix model, the divergence is somehow regularized in conjunction 
with $N$, and we obtain a finite value for $\mu$. 
 In order to calculate $\mu$ in the continuum approach, we might need 
some renormalization procedure.

Below, we summarize the results obtained for the 
nonperturbative properties of a noncritical string 
via the loop equations, 
or equivalently, the closed string field theory: 
\begin{description} 
 \item[(1)] 
  In the classical approximation ($g_s \rightarrow 0$) 
 of the loop equations, in a background of D-instantons, 
 we can obtain information regarding the D-instantons through the parameter $Q$ 
 in the instanton creation operator (\ref{eq:inst_op_naive}), 
 which is related to the constant $\varpi$ in the loop amplitude  
 (\ref{eq:loop_amplitude}). 
  However, the value of $Q$ cannot be determined within this approximation. 
 \item[(2)] 
  The treatment to all orders of the loop equations determines $Q$. 
 \item[(3)] 
  The chemical potential of the instanton 
 cannot be determined using the loop equation approach. 
\end{description} 
 It may be useful to compare these three results with the results concerning 
 D-branes obtained in critical string theory: 
\begin{description} 
 \item[\mbox{\boldmath $(1)$}] 
  The boundary state for the D-brane 
  is constructed in the closed string theory. 
  The normalization of the boundary state 
  cannot be fixed if only closed strings are considered. 
 \item[\mbox{\boldmath $(2)$}] 
  The normalization of the boundary state of the D-brane 
  is determined through comparison with the one-loop amplitude 
  of the open string. 
 \item[\mbox{\boldmath $(3)$}] 
  In order to determine the probability of the creation or annihilation of 
  D-branes, 
  which is the same as the chemical potential $\mu$, 
  we need knowledge concerning the nonperturbative dynamics of an open string 
  \cite{McGreevy:2003kb,McGreevy:2003ep,Klebanov:2003km}. 
 \end{description} 
 Although we do not know the critical string version of 
what we have done for noncritical string theory using the loop 
equation approach, 
we can see that 
there is a similarity 
between the situation for critical string theory 
and that for noncritical string theory.

\section{Conclusion} 
\label{sec:conclusion} 
\setcounter{equation}{0}
One possible  non-perturbative formulation of string theory is that which uses closed string field theory or loop equations. The situation would be simple if string theory could be formulated non-perturbatively using only closed strings. However, what is implied by the facts we have just determined is that a formulation based on only closed strings may not be able to incorporate all of the non-perturbative effects. For example, to calculate the chemical potential for the instanton, we need to know the cut-off dependence of the cylinder contribution. As we have seen, this dependence is well-defined in the matrix model, but not in the loop equations formulated in continuum variables. 

Using the loop equations in the classical limit, that is, the large-$N$ limit, one can determine the loop amplitudes in an instanton background, up to a constant factor, which is denoted by $Q$ in \S 4.  Treating the loop equations to all orders, one can further determine this factor. In terms of the D-brane of critical string theory, these two procedures, respectively, correspond to determining the boundary state up to the normalization from the boundary condition and determining the normalization of the boundary state using the one-loop calculation of the open string in the dual channel. Thus, the calculation to all orders of the loop equations is powerful enough to determine the factor that corresponds to the normalization of the boundary state, without the need for the calculation in the dual channel. Despite this fact, however, it cannot determine the chemical potential for an instanton. This corresponds to, in critical string theory, determining the weight of the annihilation and creation of D-branes from the dynamics of the tachyon.
We have shown that the matrix model does determine the chemical potential universally. Thus, at least for noncritical string theory, the closed string does not describe the nonperturbative effects completely and it seems that the description offered by the matrix model is more fundamental. If this is also the case for critical string theory, it strongly suggests that the non-perturbative formulation of string theory must contain degrees of freedom corresponding to open strings or matrices.

Because the chemical potential of the D-instanton is a universal quantity, 
it is conceivable that there exists some continuum approach to calculate it. 
Open string field theories for noncritical strings have been constructed in Refs.
\cite{Mogami:1994ip}-\cite{Nakazawa:1997wy}. It is an intriguing 
problem to calculate the value of the chemical potential using such theories. 

The calculation of the chemical potential employed in this paper is also applicable to other matrix models. The universality of the chemical potential should be checked in the case of other noncritical strings using the two-matrix model, for example. Another interesting matrix model is the two-cut model, which corresponds to super Liouville theory \cite{Douglas:2003up}. These problems are left for future studies.

In this paper, we have been attempting to determine the fundamental degrees of freedom in the nonperturbative formulation of string theory. It is suggested that we should incorporate matrix, or at least something that has endpoints, such as an open string. So far our investigation has been limited to the noncritical string theory case. 
However, it is conceivable that a similar situation exists 
for critical string theory. 
Should this issue be settled, it would be a great leap 
toward answering the ultimate question, What is string theory?

\vspace{0.5cm}
{\noindent \large \bf Acknowledgements }

\noindent The authors would like to thank M. Fukuma, T.~Matsuo, S.~Matsuura, Y. Sato, S. Sugimoto, F. Sugino, S. Yahikozawa and T. Yoneya  for fruitful discussions. 
This work is supported  in part by Grants-in-Aid for Scientific Research (13135101,13135213,13135223, 13135224, 13640308, 15740173) 
and the Grant-in-Aid for the 21st Century COE ``Center 
for Diversity and Universality in Physics" from the Ministry of Education, 
Culture, Sports, Science and Technology (MEXT) of Japan.
The work of T.K. is supported in part by a Special Postdoctoral Researchers Program.

\section*{Appendix A: \mbox{\boldmath $\phi^3$}-theory} 
\renewcommand{\theequation}{A.\arabic{equation}}
\setcounter{equation}{0}
 In this appendix, we consider one Hermitian matrix model 
with the partition function 
\begin{eqnarray} 
 Z_N 
 = 
 \int d\phi\,\exp\left(-N\,{\rm tr}\,V(\phi)\right) \, , 
  \nonumber \\ 
 V(x) = \frac{1}{2}\,x^2 - \frac{g}{3}\,x^3 \, , 
\end{eqnarray} 
to explain the convention used in the text. 
 The critical value of the coupling constant in this theory 
is 
\begin{equation} 
 g_\star = \frac{1}{2\cdot 3^{\frac{3}{4}}} \, . 
\end{equation} 
 We define the square root part of the resolvent $R(z)$ 
in the null instanton sector by 
\begin{equation} 
 \left(\widetilde{W}_0(z)\right)^2 
 \equiv 
 \frac{1}{4} 
 \left( 
  \left(V^\prime(z)\right)^2 + f_0(z) 
 \right) \, , 
\end{equation} 
where $ f_0(z) $ is a polynomial of degree $1$.
 For $g < g_\star$, 
$\left(W_0(x)\right)^2$ 
has three zeros related as $x_{(1)} < x_{(2)} < x_{(0)}$. 
 The quantity $\left(W_0(x)\right)^2$ is 
negative on the interval $(x_{(1)},\,x_{(2)})$, 
where the eigenvalues are distributed continuously 
in the large-$N$ limit. 
 At $x_{(0)}$, 
the first derivative of $\left(W_0(x)\right)^2$ vanishes . 
 For $g \rightarrow g_\star$, 
$x_{(2)}$ and $x_{(0)}$ coincide at 
\begin{equation} 
 x_\star 
 = 
 \frac{3 + \sqrt{3}}{6}\,\frac{1}{g_\star} \, , 
  \label{eq:x_star_phi3}
\end{equation} 
while $x_{(1)}$ approaches  
\begin{equation} 
 x_{(1)}(g_\star) 
 = 
 - \frac{\sqrt{3} - 1}{2}\,\frac{1}{g_\star} \, . 
  \label{eq:x1_critical_phi3}
\end{equation} 

 We would like to determine 
(some ratios of) 
the constants $\beta$, $\gamma$ and $\kappa$ 
in the double scaling limit 
(\ref{eq:double_scaling_limit}), 
so that the string equation and 
the sphere contribution to the free energy 
take the forms 
\begin{eqnarray} 
 && 
 t = \left(u(t)\right)^2 - \frac{g_s^2}{6}\,\del_t^2 u(t) 
  \, , \nonumber \\ 
 && 
 \left. F(t) \right|_{g_s \rightarrow 0} 
 = 
 - \frac{1}{g_s^2}\, \frac{4}{15}\,t^{\frac{5}{2}} \, . 
  \label{eq:string_equation_and_free_energy}
\end{eqnarray} 
 After integrating over the angular variables of $\phi$, 
$Z_0$ can be written 
\begin{eqnarray} 
 Z_0 &=& 
 \prod_{j=1}^N \int d\mu(x_j)\, 
 \left(\Delta(\{x_j\}_{j=1,\,\cdots,\,N})\right)^2 \, ,  
  \nonumber \\ 
 d\mu(x) &\equiv& dx\,e^{-N\,V(x)} \, , 
\end{eqnarray} 
where $\Delta(\{x_j\}_{j=1,\,\cdots,\,N})$ is 
the Vandermonde determinant, 
\begin{equation} 
 \Delta(\{x_j\}_{j=1,\,\cdots,\,N}) 
 = 
 \prod_{i > j} (x_i - x_j) \, . 
\end{equation} 
 From the normalization of $P_n(x)$, 
it can be written as 
\begin{equation} 
 \Delta(\{x_j\}_{j=1,\,\cdots,\,N}) 
 = 
 \det_{1 \le i,\,j \le N} P_{i-1}(x_j) \, . 
\end{equation} 
 By using this and the orthogonality of 
$\left\{P_n\right\}_{n \in {\bf Z}_+ \cup \{0\}}$, $Z_0$ becomes 
\begin{eqnarray} 
 Z_0 
 &=& 
 N! \prod_{n=1}^N h_{n-1} 
  \nonumber \\ 
 &=& 
 N! \prod_{n=1}^{N-1} r_n^{N-n} 
  \nonumber \\ 
 &=& 
 \exp 
 \left( 
  {\rm ln}\,N! 
  + 
  \sum_{n=1}^{N-1} (N-n)\,{\rm ln}\,r_n 
 \right) \, , 
\end{eqnarray} 
where $r_n$ is defined by 
\begin{equation} 
 r_n 
 \equiv 
 \frac{h_n}{h_{n-1}} \quad (n \in {\bf Z}_+) \, . 
\end{equation} 
 By subtracting ${\rm ln}\,N!$, 
the free energy $F = {\rm ln}\,Z_0$ takes the form 
\begin{equation} 
 F = \sum_{n=1}^{N-1} (N - n)\,{\rm ln}\,r_n \, . 
  \label{eq:free_energy_in_r}
\end{equation} 

 In order to compute $F$, we need to know $\{r_n\}$. 
 From the equations 
\begin{eqnarray} 
 n h_{n+1} 
 &=& 
 \int d\mu(x)\,\frac{d P_n(x)}{dx}\,P_{n-1}(x) \, ,
  \nonumber \\ 
 0 
 &=& 
 \int dx\, 
 \frac{d}{dx} 
 \left( 
  e^{-N\,V(x)} \left(P_n(x)\right)^2  
 \right) \, , 
\end{eqnarray} 
the recursion relations for $\{r_n\}$ and $\{s_n\}$ can be derived: 
\begin{eqnarray} 
 && 
 r_n \left(1 - g\left(s_n + s_{n+1}\right)\right) 
 = \frac{n}{N} \, , 
 \nonumber \\ 
 && 
 s_n^2 -\frac{1}{g}\,s_n + r_n + r_{n+1} = 0 \, . 
  \label{eq:recr_r_s}
\end{eqnarray} 
 The existence of the weak coupling limit $g \rightarrow 0$ 
selects one of the solutions of the second equation, namely 
\begin{equation} 
 s_n 
 = 
 \frac{1 - \sqrt{1 - 4 g^2(r_n + r_{n+1})}}{2 g} \, . 
  \label{eq:s_by_r}
\end{equation} 
 When $r_n$ approaches a continuous function of 
$\xi \equiv \frac{n}{N} = {\cal O}(N^0)$ 
in the large-$N$ limit, 
the first equation in (\ref{eq:recr_r_s}) implies 
that 
\begin{equation} 
 r_n 
 = 
 r\left(\frac{n}{N}\right) 
 + 
 \frac{1}{2 N^2}\,
 \frac{d^2 r}{d \xi^2}\left(\frac{n}{N}\right) 
 + {\cal O}\left(\frac{1}{N^4}\right) \, . 
  \label{eq:r_1/N}
\end{equation} 
 In the double scaling limit (\ref{eq:double_scaling_limit}), 
we scale $\xi$ and $r(\xi)$ as  
\begin{eqnarray} 
 & 
 \displaystyle{ 
  g^2 \xi = g_\star^2\, e^{- a^2 (2\beta)\,s} \, , 
 }& \nonumber \\ 
 & 
 \displaystyle{ 
  r(\xi) = \frac{1}{12 g_\star^2} \, e^{-a\,\tau\,u(s)} \, . 
 }& 
  \label{eq:scaling_for_xi_r}
\end{eqnarray} 
 By applying Eqs.~(\ref{eq:s_by_r}) and (\ref{eq:r_1/N}) 
to the first recursion equation in Eq.~(\ref{eq:recr_r_s}), 
we obtain the Painlev\'e (I) equation 
governing the singular dependence of the specific heat $u$ 
on $t$ in the limit (\ref{eq:double_scaling_limit}), 
(\ref{eq:scaling_for_xi_r}):
\begin{equation} 
 t 
 = 
   \frac{3}{4}\,\frac{\tau^2}{\beta}\, (u(t))^2 
 - g_s^2\,\frac{1}{32}\,\frac{\tau \kappa^2}{\beta^3} 
   \frac{d^2 u(t)}{dt^2} \, . 
  \label{eq:Painleve-eq}
\end{equation} 
 Also, the free energy in Eq.~(\ref{eq:free_energy_in_r}) 
becomes 
\begin{equation} 
 F(t) 
 = 
 \frac{1}{g_s^2}\, 
 \frac{4\,\beta^2 \tau}{\kappa^2} 
 \int_t^\infty ds\, (t - s)\,u(s) \, . 
  \label{eq:free_energy_dsl} 
\end{equation} 
 From Eq.~(\ref{eq:Painleve-eq}), 
the sphere contribution to $u(t)$ reads 
\begin{equation} 
 \left.u(t)\right|_{g_s \rightarrow 0} 
 = 
 \frac{2}{\sqrt{3}}\,\sqrt{\frac{\beta}{\tau^2}}\,\sqrt{t} \, .   
\end{equation} 
 By inserting this into Eq.~(\ref{eq:free_energy_dsl}), 
the universal part of the sphere contribution of $F(t)$ 
is found to be 
\begin{equation} 
 \left. F(t) \right|_{g_s \rightarrow 0} 
 = 
 - \frac{1}{g_s^2} \frac{8}{\sqrt{3}}\, 
   \frac{\beta^3}{\kappa^2 \tau}\, 
   \sqrt{\frac{\tau^2}{\beta}} 
   \times \frac{4}{15}\,t^{\frac{5}{2}}  \, . 
  \label{eq:free_energy_shpere}
\end{equation} 
 By requiring the string equation (\ref{eq:Painleve-eq}) 
to take the form given in Eq.~(\ref{eq:string_equation_and_free_energy}), 
we obtain 
\begin{eqnarray} 
 && 
 \frac{\tau^2}{\beta} = \frac{4}{3} \, , 
  \nonumber \\ 
 && 
 \frac{\kappa^2 \tau}{4 \beta^3} = \frac{4}{3} \, , 
  \label{eq:ratios_in_phi3}
\end{eqnarray} 
 We note that the same combinations of $\kappa$, $\beta$ and $\gamma$ 
appear in Eq.~(\ref{eq:free_energy_shpere}). 
 By using the values in Eq.~(\ref{eq:ratios_in_phi3}), 
it is easy to see that the sphere contribution 
necessarily takes the form given
in Eq.~(\ref{eq:string_equation_and_free_energy}). 
 The following are useful expressions derived from Eq.~(\ref{eq:ratios_in_phi3}): 
\begin{eqnarray} 
 \tau &=& 
 \frac{2}{\sqrt{3}}\,\beta^{\frac{1}{2}} \, ,
  \nonumber \\ 
 \kappa^2 
 &=& 
 \frac{8}{\sqrt{3}}\, \beta^{\frac{5}{2}} \, . 
  \label{eq:param_ratio_in_phi3}
\end{eqnarray} 

 Next, we would like to take $\frac{\beta}{\gamma^2}$ 
so that $\left(W_0(\zeta)\right)^2$ 
approaches  the conventional $\widetilde{w}_0(\zeta)$ 
in Eq.~(\ref{eq:loop_amplitude}) 
in the double scaling limit. 
 If we choose 
\begin{equation} 
 \frac{\beta}{\gamma^2} 
 = 
 \frac{3}{8}\,(2 + \sqrt{3}) 
 = 
 \frac{9}{4}\,\left(g_\star x_\star\right)^2 \, , 
  \label{eq:beta/gamma2_in_phi3}
\end{equation} 
$\left(W_0(x)\right)^2$ approaches  
\begin{equation} 
 \left(W_0(x)\right)^2 
 = 
 a^3 \times Z_W \left(w_0(\zeta)\right)^2 
 + {\cal O}(a^4) \, , 
\end{equation}  
where $Z_W$ is given by 
\begin{equation} 
 Z_W 
 = \gamma^3 \times \frac{9 + 5 \sqrt{3}}{6} 
 = 6 \gamma^3 (g_\star x_\star)^3 \, . 
  \label{eq:Z_W_phi3} 
\end{equation} 

\section*{Appendix B: \mbox{\boldmath $\phi^4$}-theory} 
\renewcommand{\theequation}{B.\arabic{equation}}
\setcounter{equation}{0}
 Here we treat the Hermitian one matrix model 
with a quartic potential that is invariant under the transformation 
$\phi \rightarrow -\phi$, 
\begin{equation} 
 V(x) = \frac{1}{2}\,x^2 - \frac{g}{4}\,x^4 \, . 
\end{equation} 
 The value of the critical coupling constant is 
\begin{equation} 
 g_\star = \frac{1}{12} \, . 
  \label{eq:g_star} 
\end{equation} 
 $\left(W_0(x)\right)^2$ in $\phi^4$-theory 
has four zeros positioned symmetrically: 
$\pm x_{(0)},\,\pm x_{(1)}$ 
($0 < x_{(1)} < x_{(0)}$). 
 For the one-cut solution, 
a cut runs along $(-x_{(1)},\,x_{(1)})$. 
 Assuming the existence of the weak coupling limit, 
$x_{(0)}$ and $x_{(1)}$ are given by 
\begin{eqnarray} 
 x_{(0)}
 &=& 
 \sqrt{\frac{1}{3 g}\left(2 + \sqrt{1 - 12 g}\right)} \, , 
  \nonumber \\ 
 x_{(1)} 
 &=& 
 \sqrt{\frac{2}{3 g}\left(1 - 12 g\right)} \, .
\end{eqnarray}
 For $g \rightarrow g_\star$, 
$x_{(0)}$ and $x_{(1)}$ come to coincide at 
\begin{equation} 
 x_\star = 2 \sqrt{2} \, . 
  \label{eq:x_star_phi4} 
\end{equation} 
 $Z_2$-symmetry implies that 
the coefficients $s_n$ 
appearing in the recursion relations for $P_n(x)$ 
vanish. 
 Thus, the recursion relations for $r_n$ are found as 
\begin{equation} 
 \frac{n}{N} 
 = r_n \left(1 - g (r_{n-1} + r_n + r_{n+1}) \right) \, . 
\end{equation} 
 We would like to obtain the form of this equation 
in the double scaling limit (\ref{eq:double_scaling_limit_phi4}), 
together with 
\begin{eqnarray} 
 g \xi &=& 
 g_\star e^{-a^2\,\beta\,s} \, , 
  \nonumber \\ 
 r(\xi) &=& 
 \frac{1}{6 g_\star}\, e^{-a\,\tau\,u(s)} \, , 
  \label{eq:beta_tau_phi4} 
\end{eqnarray} 
where $\xi \equiv \frac{n}{N}$. 

 In contrast to the $\phi^3$ potential, 
the $\phi^4$ potential has two local maxima, which are located symmetrically 
with respect to $\phi = 0$. 
 Thus, it is necessary in the double scaling limit to focus 
on both maxima simultaneously. 
 Therefore, 
we fix $\beta$, $\gamma$ and $\kappa$ in 
Eq.~(\ref{eq:double_scaling_limit_phi4}) 
and $\tau$ in (\ref{eq:beta_tau_phi4})
so that the string equation and 
the sphere contribution to the free energy 
take the forms 
\begin{eqnarray} 
 && 
 t = \left(u(t)\right)^2 - \frac{g_s^2}{6}\,\del_t^2 u(t) 
  \, , \nonumber \\ 
 && 
 \left. F(t) \right|_{g_s \rightarrow 0} 
 = 
 - \frac{1}{g_s^2}\,\frac{2 \cdot 4}{15}\,t^{\frac{5}{2}} \, . 
  \label{eq:string_equation_and_free_energy_phi4} 
\end{eqnarray} 
 Repeating the calculation given in Appendix A 
leads to 
\begin{eqnarray} 
 && 
 \frac{\tau^2}{\beta} = 1 \, , 
 \nonumber \\ 
 && 
 \frac{\tau \kappa^2}{\beta^3} = \frac{1}{2} \, . 
\end{eqnarray} 
 This implies that 
\begin{eqnarray} 
 \tau &=& \beta^{\frac{1}{2}} \, , 
  \nonumber \\ 
 \kappa^2 
 &=& 
 \frac{1}{2}\,\beta^{\frac{5}{2}} \, . 
  \label{eq:ratios_in_phi4}
\end{eqnarray} 
 We also demand that 
the continuum limit of $\left(W_0(x)\right)^2$ 
approaches $\left(\widetilde{w}_0(\zeta)\right)^2$: 
\begin{eqnarray} 
 \left(\widetilde{W}_0(x)\right)^2 
 &=& 
 a^3\,Z_W\,  
 \left(\widetilde{w}_0(\zeta)\right)^2 \, ,
  \nonumber \\ 
 \left(\widetilde{w}_0(\zeta)\right)^2 
 &=& 
 \left( \zeta + \sqrt{t} \right) 
 \left( \zeta - \frac{\sqrt{t}}{2} \right)^2 \, , 
 \label{eq:W0_in_phi4}
 \end{eqnarray}  
where
\begin{equation}
 Z_W  =  
 \gamma^3 \times \left( \frac{8}{3} \right)^2 \, . 
\end{equation}
 Such a requirement determines the ratio $\beta /\gamma^2$ as 
\begin{equation} 
 \frac{\beta}{\gamma^2} = 4 \, . 
  \label{eq:beta/gamma2_in_phi4} 
\end{equation} 

\section*{Appendix C: Behavior of $r$ and $s$}
\renewcommand{\theequation}{C.\arabic{equation}}
\setcounter{equation}{0}
 The orthogonal polynomials satisfy Schwinger-Dyson-type (SD) equations, 
\begin{eqnarray} 
 n h_{n+1} 
 &=& 
 \int dxe^{-N\,V(x)}\, \frac{dP_{n}(x)}{dx} P_{n-1}(x) \, , 
  \nonumber \\ 
 0 &=& 
 \int dx\, 
 \frac{d}{dx}\, \left( e^{-N\,V(x)} P_n(x) P_n(x) \right) \, .
\end{eqnarray} 
 The above SD equations 
can be translated into a set of recursion relations 
for the coefficients $r_n$ and $s_n$ 
appearing in Eq.~(\ref{Pzenkasiki}): 
\begin{eqnarray} 
 \left\{ 
  \begin{array}{l} 
   \displaystyle{ 
    f_n(r,\,s) = \frac{n}{N} 
   } \\ 
   \\ 
   g_n(r,\,s) = 0  
  \end{array} 
 \right. \, .
  \label{eq:eqs_for_r_s}
\end{eqnarray} 
 It is easy to see from their derivation 
that the above equations contain the quantities $r_n$ and $s_n$ 
in such a way that this set of equations is invariant under 
the substitutions 
\begin{equation} 
 \left\{ 
  \begin{array}{l} 
   r_{n + j} \leftrightarrow r_{n - j} 
   \\ 
   s_{n + j} \leftrightarrow s_{n - 1 - j} 
  \end{array} 
 \right. \, .  \label{eqn:invariancers}
\end{equation} 
When we take $n$ and $N$  to infinity, fixing their ratio as 
\begin{equation}
 0 \le \xi \equiv \frac{n}{N} \le 1 \, ,
\end{equation}
the above-mentioned fact lead us to introduce continuum functions that correspond to $r_n$ and $s_n$  
\begin{eqnarray} 
 r_n 
 &=& 
 r\left(\frac{n}{N}\right) 
 + {\cal O}\left(\frac{1}{N^2}\right) \, , 
  \nonumber \\ 
 s_n 
 &=& 
 s\left(\frac{n + \frac{1}{2}}{N}\right) 
 + {\cal O}\left(\frac{1}{N^2}\right) \, , \label{eq:rs_r()s()}
\end{eqnarray} 
with a possible $ {\cal O}(\frac{1}{N^2})$ correction.  With these definitions, the large-$N$ limit of (\ref{eq:eqs_for_r_s}) becomes consistent and well-defined up to ${\cal O}\left(\frac{1}{N}\right)$. This can be seen by considering the fact that a term invariant under (\ref{eqn:invariancers}), say  $s_{n + j} + s_{n - 1 - j} $,  yields $2s(\frac{n}{N}) +  {\cal O}\left(\frac{1}{N^2}\right)$ and no ${\cal O}\left(\frac{1}{N}\right)$ correction.
Substituting the coefficients $r_n$ and $s_n$ for the functions $r(\xi)$ and $s(\xi)$,
Eq.~(\ref{eq:rs_r()s()}) can be rewritten as 
\begin{eqnarray} 
 r_n 
 &=& 
 r(\xi) 
 + {\cal O}\left(\frac{1}{N^2}\right) \, , 
  \nonumber \\ 
 s_n 
 &=& 
 s(\xi) + \frac{\frac{\partial}{\partial \xi}s(\xi)}{2N} 
 + {\cal O}\left(\frac{1}{N^2}\right) \, ,
\end{eqnarray} 
if they are differentiable.
 Equation (\ref{eq:eqs_for_r_s}) determines $r(\xi)$ and
$s(\xi)$ at each order in the $\frac{1}{N}$-expansion. 

 It is a quite standard procedure 
to determine $r(\xi)$ and $s(\xi)$ explicitly. 
%
The large-$N$ limit of (\ref{eq:eqs_for_r_s}) can be summarized in the simple form
\begin{eqnarray}
f_\xi(r,s)&\equiv& \frac{1}{2\pi i}  \oint V'\left( z+ \frac{r}{z} +s\right) dz =\xi, \nonumber \\
g_\xi(r,s)&\equiv& \frac{1}{2\pi i}  \oint V'\left( z+ \frac{r}{z} +s\right) \frac{dz}{z} = 0, 
\label{eq:cont_eqs_for_r_and_s}
\end{eqnarray}
where the contour of the integral is taken around the origin.   As we tune the coupling constant $g$ in $V(\phi)$ to the critical value $g_\star$, $r(\xi)$ and $s(\xi)$  behave as
\begin{eqnarray}
r-r_c  &\sim  &(g_\star - g\xi)^\frac12\, , \nonumber \\
s-s_c  &\sim  &(g_\star - g\xi)^\frac12\, . \nonumber 
\end{eqnarray}
In the following, we give a more detailed description of the behavior of  $r(\xi)$ and $s(\xi)$ near the critical region by using the expression (\ref{eq:cont_eqs_for_r_and_s}).

 Equation (\ref{eq:cont_eqs_for_r_and_s}) yields directly the following relations for $f_\xi(r,s)$ and $g_\xi(r,s)$ as functions of $r$ and $s$
\begin{eqnarray}
\frac{\partial}{\partial r} f_\xi &=& \frac{\partial}{\partial s} g_\xi, \nonumber \\
\frac{\partial}{\partial s} f_\xi &=& r\frac{\partial}{\partial r} g_\xi .\label{fgcritical}
\end{eqnarray}
At the critical point, on the other hand, $f_\xi$ takes its maximum value $1$ under the constraint $g_\xi=0$. Hence the Jacobi determinant for  $f_\xi$ and  $g_\xi$ vanishes: 
\begin{equation}
\frac{\partial (f_\xi, g_\xi)}{\partial(r,s)} =0. \label{jacobianfg}
\end{equation}
 
Substituting (\ref{fgcritical}) into (\ref{jacobianfg}) yields 
\begin{equation}
\left({\frac{\partial}{\partial s} g_\xi}\right)^2-r \left({\frac{\partial}{\partial r} g_\xi}\right)^2=0,
\end{equation}
and further
\begin{equation}
{\frac{\partial}{\partial s} g_\xi}=-\sqrt{r} \left({\frac{\partial}{\partial r} g_\xi}\right),
\end{equation}
where we have taken the negative branch for the square root.
The constraint $g_\xi =0 $ yields
\begin{equation}
\frac{\frac{\partial}{\partial \xi} r}{\frac{\partial}{\partial \xi} s}=- \frac{\frac{\partial}{\partial  s} g_\xi}{\frac{\partial}{\partial r} g_\xi}.
\end{equation}
Thus, we obtain the following relation at the critical point and for $\xi =1$:
\begin{equation}
\frac{\frac{\partial}{\partial \xi}r}{\frac{\partial}{\partial \xi}s} = \sqrt{r_c}.
\end{equation}
Incorporating the above relation, the critical behavior of  $r$ and $s$ can be written in the form
\begin{eqnarray}
r&=&r_c -\frac12\alpha r_c \sqrt{1-\xi +\Delta \kern -1.5pt g } , \\
s&=&s_c - \frac12\alpha \sqrt{r_c}\sqrt{1-\xi +\Delta \kern -1.5pt g },
\end{eqnarray}
where $\Delta \kern -1.5pt g = \frac{g_\star-g}{g_\star}$ and $\alpha$ is a certain constant that might depend on the details of the model. The extra factor $\frac12$ accompanied by $\alpha$ is chosen so that (\ref{vprime}) takes a familiar form.

\section*{ Appendix D:  \mbox{\boldmath $P_n(x)$} in the oscillating region}
\renewcommand{\theequation}{D.\arabic{equation}}
\setcounter{equation}{0}
%
In this appendix, we evaluate $P_n(x)$ in the oscillatory region, applying analysis similar to that of the WKB method.
 In order to examine the continuity condition of the WKB method
in the present context, 
it is first necessary to know 
the location of the turning point(s) for a specified $x$. 
 We recall that $r(\xi)$ is a monotonically increasing function 
of $\xi$. 
 In $\phi^3$-theory, $s(\xi)$ is given by 
\begin{equation} 
 s(\xi) 
 = 
 \frac{1}{2} 
 \left( 
  \frac{1}{g} 
  - 
  \sqrt{\frac{1}{g^2} - 8\,r(\xi)} 
 \right) \, . 
\end{equation} 
 Thus, $s(\xi)$ is also an increasing function of $\xi$. 
 For a given $\xi$, 
we define $x_{(\pm)}(\xi)$ as
the zeros of the square root of $q(x,\,\xi)$: 
\begin{equation} 
 x_{(\pm)}(\xi) = s(\xi) \pm 2 \sqrt{r(\xi)} \, . 
  \label{eq:ends_of_cut} 
\end{equation} 
 The interval $\left[x_{(-)}(\xi),\,x_{(+)}(\xi)\right]$ 
is the range of $x$ 
on which $q(x,\,\xi)$ becomes purely imaginary. 
It is straightforward to see
that $x_{(+)}$ increases 
monotonically from $0$ to $x_{(2)} \equiv x_{(+)}(\xi = 1) > 0$ 
and that $x_{(-)}$ decreases 
monotonically from $0$ to $x_{(1)} \equiv x_{(-)}(\xi = 0) < 0$. 
 Such  behavior of $x_{(\pm)}$ reveals 
the following simple structure. 
If $x$ lies inside the cut $\left[x_{(1)},\,x_{(2)}\right]$, 
where the eigenvalues are distributed continuously, 
there is a $n_0(\xi)$ such that 
\begin{equation} 
 x \in 
 \left[ 
  x_{(-)}\left(\frac{n}{N}\right),\, 
  x_{(+)}\left(\frac{n}{N}\right) 
 \right] 
 \quad 
 n_0(x) \le n \le N \, . 
  \label{eq:def_of_n_0}
\end{equation} 
 If $x$ lies outside the cut $[x_{(1)},\,x_{(2)}]$, 
there is no such $n_0(x)$, 
so that $q(x,\,\xi)$ is real. 

 To summarize, we have learned that we should distinguish the two cases in which 
(a)\ $x$ lies outside the cut $[x_{(1)},\,x_{(2)}]$, 
and 
(b)\ $x$ lies inside the cut $[x_{(1)},\,x_{(2)}]$. 
 In case (a), there are no turning points, 
and therefore we only need the expression (\ref{eq:asymptotic_Pn}) 
for $P_n(x)$ 
and apply it to the estimation of $\det(x - \phi)^2$. 
 In case (b), 
there is a unique turning point 
$\xi_0(x) \equiv \frac{n_0(x)}{N}$ 
for a given $x$, 
and we need to obtain an expression 
for $P_n(x)$ for $n_0(x) \le n \le N$. 
 We next focus on this issue. 

 Let us recall the continuity formula of the WKB method. 
 As explained above, 
$\xi < \xi_0(x)$ is the classically forbidden region, 
and $\xi \le \xi_0(x)$ is the allowed region. 
 Thus, the continuity formula for two independent solutions reads 
\begin{eqnarray} 
 && 
 \frac{1}{\sqrt{\kappa(\xi)}}\, 
 \exp 
 \left( 
  \frac{1}{\hbar} 
  \int^\xi_{\xi_0} 
  \kappa(\xi^\prime)d\xi^\prime \right) 
 \Leftrightarrow 
 2\,\frac{1}{\sqrt{|\kappa(\xi)|}}\, 
 \sin 
 \left( 
    \frac{1}{\hbar} 
    \int^\xi_{\xi_0} |\kappa(\xi^\prime)| d\xi^\prime 
  + \frac{\pi}{4} 
 \right) \, , 
  \label{eq:contdec} \\ 
 && 
 \frac{1}{\sqrt{\kappa(\xi)}} 
 \exp 
 \left( 
  - \frac{1}{\hbar} 
    \int^\xi_{\xi_0} \kappa(\xi^\prime)d\xi^\prime \right) 
 \Leftrightarrow 
 \frac{1}{\sqrt{|\kappa(\xi)|}}\, 
 \sin 
 \left( 
  \frac{1}{\hbar} 
  \int^\xi_{\xi_0} |\kappa(\xi^\prime)| d\xi^\prime 
  + 
  \frac{3\pi}{4}  
 \right) \, . 
  \label{eq:continc} 
\end{eqnarray} 
 Here, 
$\kappa(\xi)$ 
is the wave number, and typically takes the form 
$\sqrt{U(\xi)-E}$ 
for an energy $E$ and potential $U(\xi)$ 
in a quantum mechanical system. 
 Note 
that there emerges a factor of $2$ 
in front of the sine 
on the right-hand side of Eq.~(\ref{eq:contdec}), 
while Eq.~(\ref{eq:continc}) 
does not possess such a factor. 
%
%
In the following, we show that we need a multiplicative factor of $2$ when we  continue the expression of $P_n(x)$  (\ref{eq:asymptotic_Pn}) analytically into the oscillatory region.
For simplicity, we consider the case $s=0$, that is, that in which the potential $V(x)$ consists only of polynomials of even power, but the following argument also applies to the case $s\neq 0$.

For a given $x$, there is a certain $n_{(0)}$ at which the argument of the square root changes sign in the expression of $P_n$ (\ref{eq:asymptotic_Pn}):
\begin{equation}
x^2 - 4r_{n_{(0)}} =0.
\end{equation}
In terms of the continuum variables $\xi=\frac{n}{N}$ and $\xi_{(0)}=\frac{n_{(0)}}{N}$, this point $\xi_{(0)}$ corresponds to the turning point in the WKB method. Up to this point (with $x^2 - 4r>0,\xi \leq \xi_{(0)} $), the leading-order behavior of $P_n$ in (\ref{eq:asymptotic_Pn}) is given by 
\begin{eqnarray}
P_n &\sim & \exp\left(N\int^\xi_0 d\xi \log \left( \frac{x}{2}+\frac12 \sqrt{x^2-4r}\right)\right) \\
 &\sim & \exp\left(-\frac{4N}{3x} \sqrt{r'(\xi_{(0)})(-\eta)^3} \right), \label{pnasym}
\end{eqnarray}
where we have introduced the new variable $\eta \equiv \xi -\xi_{(0)}$.

We postulate that near $\xi_{(0)}$, both for $\eta >0 $ and $\eta <0$, $P_n$, or the normalized polynomial $\psi_n\equiv \frac{1}{\sqrt{h_n}}P_n$, behaves as
\begin{equation}
\psi \sim  \sin\left(-N \eta^\frac32\right) \,  \mbox{ or } \, \,  e^{N(-\eta)^\frac32}. \label{postulation}
\end{equation}
Now, the quantities $\psi_n$ satisfy the recursion relation
\begin{equation}
x\psi_n= \sqrt{r_{n+1}}\psi_{n+1} + \sqrt{r_n}\psi_{n-1}. \label{rec}
\end{equation}
We  solve this recursion relation in the continuum limit near the turning point $\xi_{(0)}$ and show that 
the postulated forms given in (\ref{postulation}) are the actual forms.

First, from  (\ref{postulation}), the difference of $\psi$ behaves as
\begin{equation}
\frac{\partial}{\partial n}\psi =\frac{1}{N}\frac{\partial}{\partial \eta}\psi \sim \eta^{1/2} \psi.
\end{equation}
Thus for $\eta \sim 0$ , the higher-order differentials are suppressed. Expanding the recursion relation (\ref{rec}) both in $\frac{1}{N}$ and $\eta \sim 0$, we obtain the equation
\begin{equation}
\frac{\partial^2 \psi}{\partial \eta^2} + \frac{4N^2r'}{x^2}\eta\psi=0.
\end{equation}
This is the same equation as that which appears in the case of WKB method. Disregarding the factor $\frac{4N^2r'}{x^2}$ (or setting it to $1$) for the sake of solving the equation, we have
\begin{equation}
\frac{\partial^2 \psi}{\partial \eta^2} + \eta\psi=0.
\end{equation}
The following Airy function $\Phi(\eta)$ satisfies the above equation:
\begin{equation}
\Phi(\eta) = \frac{1}{\sqrt{\pi}i}\int_C \exp\left( \eta t +\frac13 t^3 \right).
\end{equation}
The asymptotic form of $\Phi(\eta)$ for $\eta \rightarrow -\infty$ is
\begin{equation}
\Phi(\eta)\sim \frac{1}{\eta^\frac14} \exp\left( -\frac23 (-\eta)^\frac32 \right).
\end{equation}
This form matches (\ref{pnasym}) if we set $\frac{4N^2r'}{x^2}$ to $1$.
In the other asymptotic region, $\eta \rightarrow \infty$, $\Phi$ becomes the trigonometric function
\begin{equation}
\Phi(\eta)\sim 2 \frac{1}{\eta^\frac14} \sin\left( -\frac23 (\eta)^\frac32 +\frac{\pi}{4} \right). \label{contosc}
\end{equation}
From this,  we conclude that there should be  factor of $2$ in front of the sine function, as in (\ref{contosc}),  when we continue (\ref{pnasym}) beyond the turning point.
Also, both of the above asymptotic forms are consistent with (\ref{postulation}).

\section*{Appendix E: Evaluation of \mbox{\boldmath$\left<\det (x - \phi)^2\right>$} for $x$ inside the cut and a 
simpler evaluation of 
\mbox{\boldmath$\int_{x_{(1)}}^{x_{(2)}} dx\,e^{-NV_{\rm eff}(x)}$ }
}
\renewcommand{\theequation}{E.\arabic{equation}}
\setcounter{equation}{0}
In this appendix, we present a calculation that allows for the evaluation of $\left<\det (x - \phi)^2\right>$ when $x$ lies inside the cut. We also present a simpler evaluation of the integral of $e^{-NV_{\rm eff}(x)}$ over the cut.

%

First, let us evaluate $\left<\det (x - \phi)^2\right>$ inside the cut (\ref{eq:det_truncated}):
$$
 \left<\det (x - \phi)^2\right>
 = 
 \sum_{i= n_0(x)}^N 
 \left(P_i(x)\right)^2 
 \prod_{j=0}^{N-1-i} r_{N-j} \, . 
$$
 The expression for $ \left(P_i(x)\right)^2$ is given by (\ref{Psquare}) .
 The other factor, $\prod_{j=0}^{N-1-i} r_{N-j}$, is evaluated as 
\begin{eqnarray} 
 \prod_{j=0}^{N-1-i} r_{N-j} 
 &=& 
 \exp 
 \left[ 
  \sum_{j=0}^{N-1-i} {\rm ln}\,r_{N-j} 
 \right] 
  \nonumber \\ 
 &=& 
 \exp 
 \left[ 
  N 
  \int_{1 + \frac{1}{2N}}^{\frac{n}{N} + \frac{1}{2N}} 
  d\xi\, {\rm ln}\,r(\xi) 
  + {\cal O}\left(\frac{1}{N}\right) 
 \right] 
  \nonumber \\ 
 &=& 
 \exp 
 \left[ 
  N 
  \int_{1}^{\frac{n}{N}} d\xi\, {\rm ln}\,r(\xi) 
  + 
  \frac{1}{2N}\,{\rm ln}\,r\left(1\right) 
  - 
  \frac{1}{2N}\,{\rm ln}\,r\left(\frac{i}{N}\right) 
 \right] 
  \nonumber \\ 
 && \quad 
 \times 
 \left( 
  1 + {\cal O}\left(\frac{1}{N}\right) 
 \right) \, . 
\end{eqnarray} 
 Hence, each term in Eq.~(\ref{eq:det_truncated}) becomes 
\begin{eqnarray} 
 \sum_{i= n_0(x)}^N \left(P_i(x)\right)^2 
 \prod_{j=0}^{N-1-i} r_{N-j} 
 &=&  \sum_{i= n_0(x)}^N 
2\,
 \frac{\left|k^{(0)}\left(x,\,\frac{i}{N}\right)\right|} 
      {\left|q\left(x,\,\frac{i}{N}\right)\right|} 
 \exp 
 \left[ 
 2 N\,{\rm Re}\,\int_0^{\frac{i}{N}} d\xi\,{\rm ln}\,k^{(0)}(x,\,\xi) 
 \right] 
  \nonumber \\ 
 && \quad 
 \times 
 \frac{\sqrt{r\left(1\right)}} 
      {\sqrt{r\left(\frac{i}{N}\right)}} 
 \exp 
 \left[ 
  N \int_{\frac{i}{N}}^{1} d\xi'\, 
  {\rm ln}\,r(\xi') 
 \right] 
 \left( 
  1 + {\cal O}\left(\frac{1}{N}\right)
 \right)
  \nonumber \\ 
 &=& 
N\int_{\frac{n_{(0)}}{N}}^1 d\xi \frac{2\sqrt{r\left(1\right)}} 
      { \left|q\left(x,\,\xi\right)\right|}\, 
\nonumber \\ && \quad \times \exp 
 \left[ 
2 N\int_0^{\frac{n_{(0)}}{N}} d\xi\,{\rm ln}\,k^{(0)}(x,\,\xi)   + N 
  \int_{n_{(0)}}^{1} d\xi\, {\rm ln}\,r(\xi) 
 \right] 
  \nonumber \\ 
 && \quad 
 \times 
 \left( 
  1 + {\cal O}\left(\frac{1}{N}\right)
 \right) \, ,  \label{eq:detxphiincut}
\end{eqnarray} 
where we have used 
$\left|k^{(0)}\left(x,\,\xi\right)\right|^2 = r(\xi)$ when $\xi$ is greater than $\frac{n_{(0)}}{N}$.
 Now, we claim the relation
\begin{equation}
 \pi \rho(x) 
 = 
 \int_{\frac{n_0(x)}{N}}^1 d\xi\, 
 \frac{1}{\left|q\left(x,\,\xi\right)\right|} 
 \left( 1 + {\cal O}\left(\frac{1}{N}\right) \right)\, . 
\end{equation} 
 This is proved as follows. 
 Because $P_N(x) = \left<\det(x - \phi)\right>$, we have 
\begin{eqnarray} 
 \left< 
  \exp\left[ {\rm tr}\,{\rm ln}\,(x - \phi) \right] 
 \right> 
 &=& \left< \det(x - \phi) \right> 
 = P_N(x) 
  \nonumber \\ 
 &=& 
 \frac{k^{(0)}(x,\,1)}{q(x,\,1)} 
 \exp 
 \left[ 
  N \int_0^1 d\xi\, {\rm ln}\,k^{(0)}(x,\,\xi) 
 \right]  \nonumber \\ && \quad \times
 \left( 1 + {\cal O}\left(\frac{1}{N}\right) \right) \, . 
  \nonumber\\
  \label{eq:res_diff_PN}
\end{eqnarray}  
 Differentiating both sides of this equation 
with respect to $x$, we obtain 
\begin{eqnarray} 
 && 
 \left< 
  \frac{1}{N}\,{\rm tr} \left( \frac{1}{x - \phi} \right)\, 
  \det (x - \phi) 
 \right> 
  \nonumber \\ 
 && \qquad 
 = 
 \left< \det(x - \phi) \right> 
 \left( 
  \int_0^1 d\xi\,\del_x {\rm ln}\,k^{(0)}(x,\,\xi) 
  + 
  \frac{1}{N}\, 
  \del_x 
  {\rm ln} \left(\frac{k^{(0)}(x,\,\xi)}{q(x,\,\xi)}\right) 
 \right) \, . 
\end{eqnarray} 
 In the large-$N$ limit, 
the average on the left-hand side factorizes. 
 Hence, Eq.~(\ref{eq:res_diff_PN}) 
reduces at leading-order in $N$ to 
\begin{equation} 
\left< \frac{1}{N}\, 
 {\rm tr}\,\frac{1}{x - \phi} 
 \right>
 = 
 \int_0^1 d\xi\, 
 \del_x {\rm ln}\,k^{(0)}(x,\,\xi) 
  \left( 
   1 + {\cal O}\left(\frac{1}{N}\right)
  \right) \, . 
\end{equation} 
 Here we recall that 
\begin{equation} 
 \del_x {\rm ln}\,k^{(0)}(x,\,\xi) 
 = 
 \frac{1}{q(x,\,\xi)} \, . 
\end{equation} 
 Thus, the imaginary part of this relation gives 
\begin{eqnarray} 
 - 2 \pi i \rho(x) 
 &=& 
 \left< 
  \frac{1}{N}\,{\rm tr}\,\frac{1}{x + i \epsilon - \phi} 
 \right> 
 - 
 \left< 
  \frac{1}{N}\,{\rm tr}\,\frac{1}{x - i \epsilon - \phi} 
 \right> 
  \nonumber \\ 
 &=& 
 \int_0^1 d\xi 
 \left( 
    \frac{1}{q(x + i \epsilon,\,\xi)} 
  - \frac{1}{q(x - i \epsilon,\,\xi)} 
 \right) \, , 
\end{eqnarray} 
from which we obtain 
\begin{equation} 
 \rho(x) 
 = 
 \frac{i}{2\pi} 
 \int_0^1 d\xi 
 \left( 
    \frac{1}{q(x + i \epsilon,\,\xi)} 
  - \frac{1}{q(x - i \epsilon,\,\xi)} 
 \right) \, . 
  \label{eq:rho_u_diff}
\end{equation}  
 From the monotonic behavior of $x_{(\pm)}(\xi)$, 
we have seen that there is such an $n_0(x)$ that  
\begin{eqnarray} 
 && 
 q(x + i \epsilon,\,\xi) = q(x - i \epsilon,\,\xi) 
  \, ,
 \quad \left(0 \le \xi < \frac{n_0(x)}{N}\right) 
 \nonumber \\ 
 && 
 q(x \pm i \epsilon,\,\xi) 
 = \pm i \left| q(x,\,\xi) \right| 
 \, . 
 \quad \left(\frac{n_0(x)}{N} \le \xi \le 1 \right) 
\end{eqnarray} 
 Inserting this into Eq.~(\ref{eq:rho_u_diff}), 
we obtain the relation in question,
\begin{equation} 
 \rho(x) 
 = 
 \frac{1}{\pi} 
 \int_{\frac{n_0(x)}{N}}^1 d\xi\, 
  \frac{1}{\left|q(x,\,\xi)\right|} 
  \left( 1 + {\cal O}\left(\frac{1}{N}\right) \right) \, . 
\end{equation} 
Using the above relation in Eq. (\ref{eq:detxphiincut}) and introducing $\xi_0(x)=\frac{n_0(x)}{N}$, we obtain Eq. (\ref{eqn:detxphiaveraged}).

Next, we show that there is a much simpler way to compute 
the left-hand side of (\ref{zden}) , 
\begin{equation}
Z_{\rm den}\equiv\int_{x_{(1)}}^{x_{(2)}} dx\,e^{-NV_{\rm eff}(x)}
\label{Zden}
\end{equation}
directly, without using the explicit form of $V_{\rm eff}(x)$.

The point here is that we can replace the interval 
of the integration appropriately with a relative error 
within ${\cal O}(1/N)$. First, as we saw in 
\S\ref{subsec:evalofchar.poly.}, the integrand is ${\cal O}(1)$ 
inside the cut, while it decays exponentially as 
$\sim \exp (-Nf(x))$. Therefore, we can replace (\ref{Zden}) 
as
\begin{equation}
Z_{\rm den}=\int_{-\infty}^{\infty} dx\,e^{-NV_{\rm eff}(x)}, 
\end{equation}
because the contribution from outside the cut 
gives only $\sim \int_0^{\infty}\exp (-Ncx)\sim {\cal O}(1/N)$, 
where $c$ is a certain constant of ${\cal O}(1)$. 
Then, using (\ref{eqn:orthogonality}),(\ref{eqn:hratio}) 
and (\ref{eqn:DN}), we obtain 
\begin{equation}
Z_{\rm den}=(N+1)h_N\sim Nh_N.
\end{equation}

Next, we note that from (\ref{eqn:orthogonality}), $h_N$ 
is obtained as 
\begin{equation}
h_N=\int_{-\infty}^{\infty}P_N(x)^2e^{-NV(x)}dx. 
\end{equation}
However, we can again replace the interval of this integration 
`inversely' with 
\begin{equation}
h_N=\int_{x_{(1)}}^{x_{(2)}}P_N(x)^2e^{-NV(x)}dx, 
\end{equation}
up to a relative error of {\cal O}(1/N), 
for the same reason as above.
Now, using (\ref{Psquare}) with $n=N$ 
and noting that 
\begin{equation}
V^{(0)}_{\rm eff}(x)
=2\,{\rm Re}\,\int_0^1 d\xi\,{\rm ln}\,k^{(0)}(x,\,\xi)
-V(x)
\end{equation}
is constant inside the cut, as we have seen 
in \S\ref{subsec:Sinst}, we obtain 
\begin{equation}
h_N=2\sqrt{r(1)}e^{-NV^{(0)}_{\rm eff}(x_{(2)})}
     \int_{x_{(1)}}^{x_{(2)}}
     \frac{1}{|q(x,1)|}dx.
\label{eqn:h_N}    
\end{equation}
 
 Finally, let us compute the remaining integral, 
\begin{equation}
I=\int_{x_{(1)}}^{x_{(2)}}
    \frac{1}{|q(x,1)|}dx.
\end{equation}
As mentioned in \S\S\ref{subsec:asmbehP} and 
\ref{subsec:evalofchar.poly.}, the cut is the region 
for which $q(x,\xi)^2\le 0$, and, in particular, $q(x,1)$ provides 
the largest cut, $x_{(1)}<x<x_{(2)}$.  Therefore, 
$q(x,1)=\sqrt{(x-x_{(1)})(x-x_{(2)})}$, and 
\begin{equation}
I=\int_{x_{(1)}}^{x_{(2)}}
    \frac{1}{\sqrt{(x-x_{(1)})(x_{(2)}-x)}}dx=\pi.
\end{equation}
Substituting this into (\ref{eqn:h_N}), we obtain 
\begin{equation}
h_N=2\pi\sqrt{r(1)}e^{-NV^{(0)}_{\rm eff}(x_{(2)})}
\end{equation}
and 
\begin{equation}
Z_{\rm den}=2N\pi \sqrt{r(1)}e^{-NV^{(0)}_{\rm eff}(x_{(2)})},
\end{equation}
which of course agrees with the result in (\ref{zden}).

\section*{Appendix F: Chemical potential for 
\mbox{\boldmath$\phi^3$} and \mbox{\boldmath$\phi^4$} potentials}
\renewcommand{\theequation}{F.\arabic{equation}}
\setcounter{equation}{0}
In this appendix we present explicit calculations of the chemical potential for  the $\phi^3$ and $\phi^4$ potential cases. These two cases yield the same result as the other. This reveals the universality of the chemical potential.

\subsubsection*{The $\phi^3$-potential}

In this case, we use the potential 
$V(x) = \frac{1}{2}x^2 -\frac{g}{3}x^3$. 
The resolvent is written 
\begin{equation}
 R(x) = \frac{1}{2}\left[V'(x)
                    +\left(gx-gs(1)+1\right)\sqrt{(x-s(1))^2-4r(1)}\right].
\end{equation}
The chemical potential $\mu$ is given by \eqref{eqn:chempot}. 
Hence, using \eqref{veffout2} and \eqref{veffin2}, we obtain
\begin{align}
 &\mu= \frac{Z_N^{(1-\text{inst})}}{Z_N^{(0-\text{inst})}} \notag\\
  &= \frac{i}{\sqrt{\pi N}}
  \left(\frac{4}{(1-2gs(1))^2-4g^2r(1)}\right)^{\frac{1}{4}} \notag\\
  &\qquad \times\frac{\left((1-2gs(1))+\sqrt{(1-2gs(1))^2-4g^2r(1)}\right)^2}
  {8\sqrt{r(1)}\left((1-2gs(1))^2-4g^2r(1)\right)}
 \exp\left[-NV_{\text{eff}}^{(0)}(x_{(0)})
 +NV_{\text{eff}}^{(0)}(x_{(2)})\right].
\end{align}
Here, we have used the saddle point method around 
$x_{(0)}=\frac{1}{g}-s(1)$ for the integration in the numerator 
of \eqref{eqn:chempot}.

We introduce $t$ so that  the free energy can be expressed in the universal form
\begin{equation}
 F = -N^2 \frac{4}{15}\sqrt{\frac{2}{3}}
  \left(\frac{g_c^2-g^2}{g^2_c}\right)^{\frac{5}{2}}
  = - \frac{4}{15}t^{\frac{5}{2}}.
\end{equation}
Then, the chemical potential $\mu$ can be written in terms of $t$ as
\begin{equation}
 \mu = \frac{Z_N^{(1-\text{inst})}}{Z_N^{(0-\text{inst})}}
  = \frac{i}{8\cdot 3^{3/4}\sqrt \pi t^{\frac{5}{8}}}
  \ e^{-\frac{8\sqrt 3}{5}t^{\frac{5}{4}}}.
\end{equation}

\subsubsection*{The $\phi^4$-potential}

In this case, the potential is given by
$V(x) = \frac{1}{2}x^2 -\frac{g}{4}x^4$.
The resolvent is written 
\begin{equation}
 R(x) = \frac{1}{2}\left[V'(x)
                   + \left(gx^2+2gr(1)-1\right)\sqrt{x^2-4r(1)}\right].
\end{equation}
Using \eqref{veffout2}, \eqref{veffin2} and \eqref{eqn:chempot}, 
we obtain
\begin{align}
 &\mu = \frac{Z_N^{(1-\text{inst})}}{Z_N^{(0-\text{inst})}}\notag \\
  &= \frac{i}{\sqrt{\pi N}}
 \left(\frac{1}{(1-2gr(1))(1-6gr(1))}\right)^{\frac{1}{4}} \notag\\
 &\qquad\times\frac{\left(\sqrt{1-2gr(1)}+\sqrt{1-6gr(1)}\right)^2}
 {8\sqrt{r(1)} (1-6gr(1))}
 \exp\left[
      -NV_{\text{eff}}^{(0)}(x_{(0)})+NV_{\text{eff}}^{(0)}(x_{(2)})
     \right] ,
\end{align}
where $x_{(0)}=\sqrt{\frac{1}{g}-2r(1)}$.

We introduce $t$ so that  the free energy can be expressed in the universal form
\begin{equation}
 F = -N^2 \frac{4}{15}\left(\frac{g_c - g}{g_c}\right)^{\frac{5}{2}}
  = - 2\frac{4}{15}t^{\frac{5}{2}}.
\end{equation}
Here, the factor 2 reflects the fact that there are two critical
points at $x=x_\star$ and $x=-x_\star$, because the potential is even.
Then, we obtain the following value for  the chemical potential:
\begin{equation}
 \mu = \frac{Z_N^{(1-\text{inst})}}{Z_N^{(0-\text{inst})}}
  = \frac{i}{8\cdot 3^{3/4}\sqrt \pi t^{\frac{5}{8}}}
  \ e^{-\frac{8\sqrt 3}{5}t^{\frac{5}{4}}}.
\end{equation}



\end{document}